\documentclass[12pt]{article}

\usepackage{fullpage}
\usepackage{amsmath}
\usepackage{graphicx}
\usepackage{amssymb}

\numberwithin{equation}{section}

\newlength{\dummysp}
\newcommand{\tr}{\mathop{{\hbox{Tr} \, }}\nolimits}

\newcommand{\half}{\frac{1}{2}}

\newcommand{\beq}{\begin{eqnarray}}
\newcommand{\eeq}{\end{eqnarray}}
\newcommand{\nnn}{ \nonumber \\ }
\newcommand{\p}{{\partial}}

\newcommand{\e}{{\epsilon}}

\newcommand{\vev}[1]{{\langle #1 \rangle}}

\newcommand{\ord}[1]{{{\cal O}(#1)}}
\newcommand{\gappeq}{\mathrel{\rlap {\raise.5ex\hbox{$>$}}
{\lower.5ex\hbox{$\sim$}}}}
\newcommand{\lappeq}{\mathrel{\rlap{\raise.5ex\hbox{$<$}}
{\lower.5ex\hbox{$\sim$}}}}
\newcommand{\myref}[1]{(\ref{#1})}

\newcommand{\ben}{\begin{enumerate}}
\newcommand{\een}{\end{enumerate}}

\newcommand{\psib}{{\bar \psi}}

\newcommand{\ddd}{\nnn &&}

\newcommand{\bit}{\begin{itemize}}
\newcommand{\eit}{\end{itemize}}
\newcommand{\Lcal}{{\cal L}}

\newcommand{\Ocal}{{\cal O}}

\newcommand{\sss}[1]{\subsubsection{#1}}

\def\[{\left [}
\def\]{\right ]}
\def\({\left (}
\def\){\right )}

\def\nott#1{\setbox0=\hbox{$#1$}                
   \dimen0=\wd0                                 
   \setbox1=\hbox{/} \dimen1=\wd1               
   \ifdim\dimen0>\dimen1                        
      \rlap{\hbox to \dimen0{\hfil/\hfil}}      
      #1                                        
   \else                                        
      \rlap{\hbox to \dimen1{\hfil$#1$\hfil}}   
      /                                         
   \fi}                                         %

\begin{document}

\begin{titlepage}

\begin{center}
{\bf \large Anomalous dimensions on the lattice}
\end{center}

\bigskip

\bigskip

\begin{center}
Joel Giedt \\
{\it Department of Physics, Applied Physics and Astronomy \\
Rensselaer Polytechnic Institute, 110 8th Street, Troy, NY 12180 USA }
\end{center}

\bigskip

\bigskip

\begin{abstract}
We review methods and results for extracting the anomalous dimensions of operators from
lattice field theory calculations.  The most important application is the anomalous mass dimension in conformal
or nearly conformal gauge field theories which might be related to
dynamical electroweak symmetry breaking.  Some discussion of the underlying theory
of renormalization and mixing of operators is also included.
\end{abstract}

\end{titlepage}

\tableofcontents

\section{Introduction}
The fact that composite operators in an interacting
quantum field theory do not have the naive, classical scaling
dimension associated with the Gaussian fixed point, but actually have
a quantum mechanical dimension, which includes a so-called ``anomalous''
part, is well known \cite{Wilson:1970pq,Wilson:1970wp}.  (Anomalous dimensions were
anticipated in earlier work on critical phenomena, such as \cite{Patas66}.)
In some cases, the interacting theory is solvable
and one can compute these dimensions exactly.  However, that is rarely
the case.  In other cases, supersymmetry and conformal symmetry are present and the superconformal
algebra relates the scaling dimension to the charge under the $U(1)_R$
symmetry, so that at least in the case of ``chiral'' operators\footnote{A
chiral operator $\Ocal$ is one that in superspace is annihilated by
the ``dotted'' superspace covariant derivative,
$\overline{D}^{\dot\alpha} \Ocal=0$.} 
the anomalous dimension is known exactly.  Another
avenue is that the theory is weakly coupled, so that one can compute the
anomalous dimension in a perturbative approximation; this would
be true, for instance, in the case of a Banks-Zaks fixed point \cite{Banks:1981nn}.  Nonperturbative
approaches must be used when the theory is strongly coupled, as in an
asymptotically free theory that is far away from the Banks-Zaks limit.  One can use truncated Schwinger-Dyson equations
to derive estimates of the anomalous dimension, but this is an uncontrolled
approximation. Ultimately one would like to obtain the nonperturbative
estimate to arbitrary accuracy using a first principles approach with
a controlled approximation that can be systematically improved.  Lattice
gauge theory provides a tool to obtain such an estimate, but it is
fraught with technical and practical difficulties, as will be made clear
in this report.  Nevertheless, it is a route that is worth pursuing
in the case where exact methods are unavailable and the theory is
strongly interacting.

Several groups have recently computed the anomalous mass dimension,
essentially the scaling dimension of the scalar fermion bilinear operator $\Ocal_S = \psib \psi$,
in theories that are cousins of massless QCD:  fermions coupled to a nonabelian
gauge field, in the chiral limit (where the current ``quark'' mass is zero).
The interest in this topic has been motivated by the desire to
find viable models of walking technicolor (TC), where a large anomalous
mass dimension leads to condensate enhancement, 
\beq
\langle \psib \psi \rangle_{\text{ETC}} \sim \( \frac{\Lambda_{\text{ETC}}}{\Lambda_{\text{TC}}}
\)^{\gamma_*} \langle \psib \psi \rangle_{\text{TC}}
\eeq
which is important
for the suppression of flavor-changing neutral currents.  
These
flavor changing processes are mediated by the exchange of extended
technicolor (ETC) gauge bosons, and the mass scale of these particles
needs to be high in order to avoid constraints from experimental
flavor physics.  On the other hand, such a high scale would also
naively suppress the fermion masses of the standard model.  This is
avoided if the technicolor condensate that feeds into the fermion
masses is enhanced, due to a large anomalous dimension---an explicit
mathematical formulation of this will be given below.  In practice
one desires $\gamma_* \approx 1$ for viable phenomenology \cite{Chivukula:2010tn}. Thus the
lattice community has been examining the anomalous mass dimension
in various theories that might serve as walking technicolor candidates,
as well as theories that are more QCD like in order to draw a contrast.\footnote{General reviews of these lattice
studies of models relevant to dynamical electroweak symmetry breaking are given in \cite{Kuti:2014epa,DelDebbio:2014ata,DeGrand:2015zxa}.
A review focused on the dynamical generation of scale, and finite temperature transitions as a function
of the number of fermion flavors can be found in \cite{Lombardo:2014mda}.}
Several techniques for doing this have been exploited and developed.
In this report we will describe most of the methods that have been
used to date, and will attempt to compare them in terms of their
reliability and efficacy.  

The walking technicolor motivation for such studies is under stress from
the recent discovery of the Higgs boson, which appears to have all of the
properties of the minimal elementary scalar field model.  In fact, even
two Higgs doublet models such as appear in supersymmetric extensions
seem to be forced into the decoupling limit where the CP odd Higgs scalar
$A^0$ is very heavy.  Naively one might think that the lightest scalar
in a technicolor theory should be heavy.  After all in QCD the $\sigma$
resonance is of order 500 MeV, which is quite a bit larger than the
pion decay constant $f_\pi \sim 90$ MeV.  In the technicolor model,
the pion decay constant is scaled up to the Higgs vacuum expectation
value, $v = f_\pi = 246$ GeV, so one would expect the analogue of the
$\sigma$ to be of order 1 TeV, not the 125 GeV of the observed Higgs boson.
However, it has recently been pointed out that electroweak corrections,
principally the top quark loop, have the right sign and magnitude to bring
the mass of the $\sigma$ down significantly, possibly even to 125 GeV 
\cite{Foadi:2012bb},
if the $\sigma$ is perhaps a bit lighter than scaled up QCD would
predict.  A somewhat light $\sigma$ may happen in walking technicolor because the dynamics 
is significantly different and there is an approximate zero of the
$\beta$ function at some scale, which some have argued can suppress
the mass of the lightest scalar (the so-called techni-dilaton).

The lattice studies have for the most part found $\gamma \lappeq 0.5$, which is too
small for the walking scenario.\footnote{The exception seems to be theories that
are right on the lower edge of the conformal window---but this needs more study.}  However, there is the potential to solve this
problem with the four-fermion terms that will be induced by extended technicolor \cite{Fukano:2010yv}.
This may also be used to push a theory out of the conformal window, so that it really is
walking.  For this reason, detailed lattice studies of theories with four-fermion
terms in the action need to be pursued in the future.  This is not a simple matter
since introducing four-fermion terms often destroys the positivity of the fermion
measure, leading to an apparently insurmountable sign problem.\footnote{As we said in
an earlier paragraph, the lattice approach is fraught with difficulties; the sign
problem is one of them.}  Nevertheless, a judicious
choice of theory and lattice discretization can avoid this problem and would lead to
very interesting results if studied in light of the proposal implied by \cite{Fukano:2010yv}.

Apart from technicolor type models, anomalous dimensions and conformal fixed points are
important problems for a variety of reasons.  In some supersymmetric theories the
hidden sector may be nearly conformal and the anomalous dimension of some operators
can determine the impact of the hidden sector on soft parameters; this can be
significant.  Simply knowing the sign of the anomalous dimension of some operators
can give answers to qualitative questions.  Of course one would also have to solve
the supersymmetry problem on the lattice in order to conduct such a study.\footnote{Another difficulty.}

Knowledge of the running of the anomalous dimension in asymptotically free gauge theories
has also played a role in holographic calculations \cite{Erdmenger:2014fxa}.  In that
context the running of $\gamma$ translates into a mass-squared for the scalar
dual to ${\bar q} q$ that depends on the AdS radius, $m^2 = m^2(r)$.  Here it
is interesting that the Breitenlohner-Freedman bound occurs precisely when $\gamma=1$,
the number which is sought after in the lattice gauge theory studies.  Such a value
may indeed occur near the critical number of flavors, such as $N_c=3$ with $N_f=10$
fundamental representation fermions \cite{Appelquist:2012nz}, or $N_c=2$ with $N_f=1$ adjoint Dirac 
fermions \cite{Athenodorou:2014eua}.

\section{Anomalous dimensions}
Correlation functions of the bare fields $\phi_0$ and bare operators $\Ocal_0$
require renormalization in order to obtain finite answers:\footnote{This formula
ignores mixing of operators; we will turn to this matter shortly.}
\beq
\langle 0 | \phi(x_1) \cdots \phi(x_n) \Ocal(y) | 0 \rangle
= Z_\Ocal Z_\phi^{-n/2} \langle 0 | \phi_0(x_1) \cdots \phi_0(x_n) \Ocal_0(y) | 0 \rangle
\eeq
That this even works to render the left-hand side finite is in itself amazing:  an
infinite number of correlation functions (taking all possible choices of $n$) are
renormalized just using two renormalization constants $Z_\phi$ and $Z_\Ocal$ (of course
the bare masses and couplings will also need to be adjusted to cancel the infinities,
but for a renormalizable theory this is a finite set).  
The $Z$ factors depend on the renormalization scale $\mu$.  In particular,
\beq
\Ocal(\mu;y) = Z_\Ocal(\mu) \Ocal_0(y)
\label{Zren}
\eeq
The $Z$ factor for the operator satisfies a renormalization group equation,
\beq
\gamma_\Ocal(g(\mu)) = -\frac{d}{d \ln \mu} \ln Z_\Ocal(\mu)
\label{ganomi}
\eeq
where we show explicitly that the anomalous dimension $\gamma_\Ocal$ is a function of the
running coupling $g(\mu)$.  Since the bare operator is independent of the renormalization
scale $\mu$,
\beq
\frac{d}{d \ln \mu} \Ocal(\mu;y) = -\gamma_\Ocal \Ocal(\mu;y)
\eeq

These equations ignore the issue of mixing, which is often not negligible.  To take this
into account, we generalize to a bare operator basis $\Ocal_{0,i}$ and a renormalized
set of operators $\Ocal_i$.  Then these are related by
\beq
\Ocal_i(\mu;y) = Z_{ij}(\mu) \Ocal_{0,j}(y)
\eeq
where of course there is a sum over $j$ on the right-hand side.  The renormalization
group equation generalizes to
\beq
\frac{d}{d \ln \mu} Z_{ij}(\mu) = -\gamma_{ik}(g(\mu)) Z_{kj}(\mu)
\label{gawmi}
\eeq
Contracting with the bare operators, we see then that there is a corresponding flow
in the space of renormalized operators,
\beq
\frac{d}{d \ln \mu} \Ocal_{i}(\mu;y) = -\gamma_{ik}(g(\mu)) \Ocal_{k}(\mu;y)
\eeq

In the case of the fermion mass operator $\psib\psi$, since $m \psib\psi$ is RG
invariant\footnote{A physical quantity $P$ is RG invariant if it satisfies
$$ \mu \frac{d}{d\mu} P = \( \mu \frac{\p}{\p \mu} + \beta(g) \frac{\p}{\p g}
+ \gamma(g) m \frac{\p}{\p m} \) P = 0 $$
I.e., it is a solution to the Callan-Symanzik equation.} 
(it appears in the trace of the energy momentum tensor, a conserved
current), the anomalous dimension of this operator is also related to the dimension
of the mass:
\beq
\gamma_{\psib\psi} = \gamma_m = - \frac{d \ln m}{d \ln \mu}
\eeq

The operators will have a canonical, or engineering, dimension
\beq
[ \Ocal_i ] = d_i
\eeq
Up to $\ord{a}$ terms, the operators
will only mix with those of the same dimension or lower.
This then takes the form
\beq
\Ocal_i(\mu;x) = \sum_{j \in S} a^{-d_i+d_j} C_{ij}(\mu a) \Ocal_{0,j}(x) + \ord{a}
\eeq
where $S$ is the set of operators with $d_j \leq d_i$.  The operators
of dimension $d_j < d_i$ do not affect the anomalous dimensions of
the operators of dimension $d_i$ \cite{Testa:1998ez}.  Nevertheless, they play
an important role, for instance in the determination of counterterms to restore symmetries;
see for example \cite{Testa:1998ez} for a discussion of axial Ward identities
with Wilson fermions, or \cite{Farchioni:2001wx} for supersymmetry Ward identities
again with Wilson fermions.  The $\ord{a}$ terms that are not shown are
important in $\ord{a}$ improvement.  It is quite expensive to reduce the
lattice spacing in a lattice simulation because the spatial size $L=Na$ must
be kept constant to avoid finite size effects, and the number of degrees
of freedom will increase as $1/a^4$.  In addition, as one reduces the lattice
spacing one simulates closer to a second order critical point and so there is
a critical slowing down in the algorithms.  In practice it is better to first
reduce the discretization errors as much as possible.  This can be achieved
by fine-tuning irrelevant operators in the action and by making $\ord{a}$
improvements to operators that are being measured.

The anomalous dimensions of composite operators also make an appearance in
a generalization of the Callan-Symanzik equations for renormalized Green functions,
\beq
\[ \mu \frac{\p}{\p\mu} + \beta(g) \frac{\p}{\p g} + n \gamma(g) + m \gamma_\Ocal(g) \]
\langle 0 | \Phi^R(x_1) \cdots \Phi^R(x_n) \Ocal^R(y_1) \cdots \Ocal^R(y_m) | 0 \rangle =0
\eeq
Here $\Phi^R$ represent the renormalized elementary fields.
This also provides a rubric for finding the anomalous dimensions from the
counterterms $\delta Z_\Ocal$ which are involved in rendering the Green
function
\beq
G^{R,n,m}(x_1,\ldots,x_n;y_1,\ldots,y_m) =
\langle 0 | \Phi^R(x_1) \cdots \Phi^R(x_n) \Ocal^R(y_1) \cdots \Ocal^R(y_m) | 0 \rangle
\eeq
finite.  Explicit examples will be considered below.

\section{Non-lattice methods}
\subsection{Perturbation theory}
In a theory of only fermions with gauge interactions, the one loop anomalous mass dimension is universal
(scheme independent) and given by
\beq
\left. \gamma_{\psib\psi} \right|_{\text{1-loop}} = -\frac{3 C_2(R) \alpha}{2\pi}
\label{unigam}
\eeq
The minus sign originates from our definition \myref{ganomi} with Z being used to obtain
the renormalized operator from the bare operator as in \myref{Zren}, rather than
the other way around.
Universality is easily seen, since it would correspond to a redefinition
$\alpha \to \alpha + \ord{\alpha^2}$, which will not affect the
one-loop result.

The two-loop contribution in the MS or $\overline{\text{MS}}$ schemes is given by
\beq
\left. \gamma_{\psib\psi} \right|_{\text{2-loop}} = -\frac{C_2(R) \alpha^2}{16 \pi^2}
\( \frac{203}{6} N_c - \frac{3}{2 N_c} - \frac{10}{3} N_f \)
\eeq
Also in these schemes the three and four loop results are known and given in 
Refs.~\cite{Chetyrkin:1997dh,Vermaseren:1997fq}.

\subsection{Unitarity bounds}
Unitarity constraints on the conformal group \cite{Mack:1975je} imply that $\gamma_* \leq 2$.
For supersymmetric theories, further constraints
exist, as studied extensively in \cite{Minwalla:1997ka}
and reviewed in \cite{Wiegandt:2012eu}.  Thus in general
we can use the power of the conformal group, or yet more power of the superconformal group,
to extract information about the scaling dimensions.  Of course, this requires that the
theory be conformal.  A necessary, but not always sufficient condition, is that it
be scale invariant.  However, as pointed out in \cite{Minwalla:1997ka}, a quantum field
theory must include a cutoff, and a scale transformation scales the cutoff.  So if
there is any cutoff dependence in the quantum theory, say $\Lambda dg/d\Lambda = \beta(g) \not= 0$,
then the theory will not be conformally invariant.\footnote{Note that in this
argument, $g$ represents the bare coupling, and the IR coupling is being held
fixed as the cutoff scale is changed.  Thus in the case where there is
some flow, the bare coupling becomes cutoff dependent.  Also note that
the $\beta$ function here is the bare $\beta$ function.  This distinction
is familiar to lattice gauge theorists.}  Hence the need for a vanishing
$\beta$ function.

\subsection{Example:  QED at one loop}
Here, we will work in the limit of vanishing electron mass.  As will be seen, this also
corresponds to a mass independent scheme for subtracting the infinities.  

\begin{figure}
\begin{center}
\begin{tabular}{cc}
\includegraphics[width=2in]{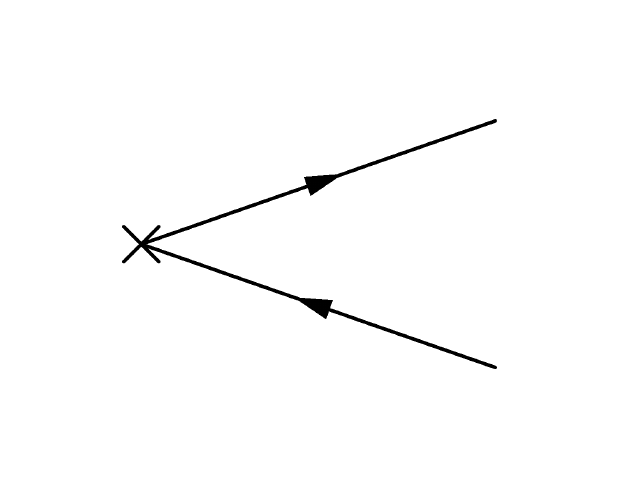} &
\includegraphics[width=2in]{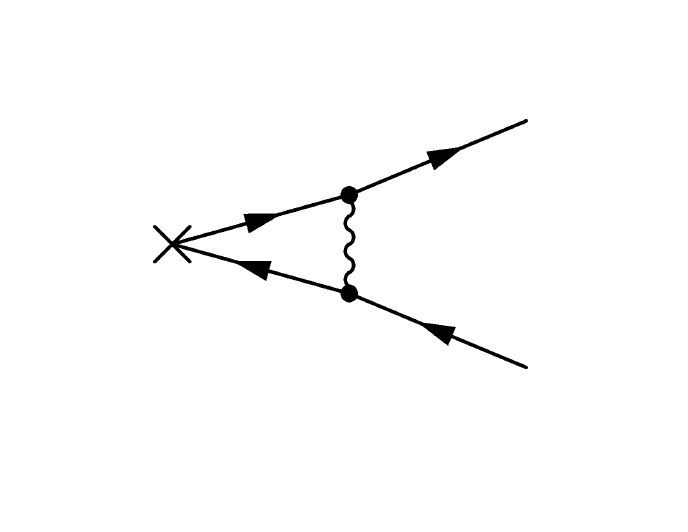} \\
\includegraphics[width=2in]{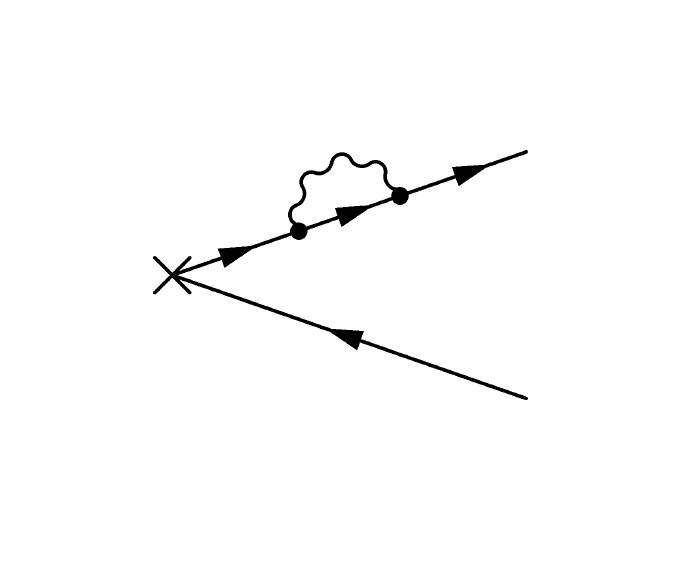} &
\includegraphics[width=2in]{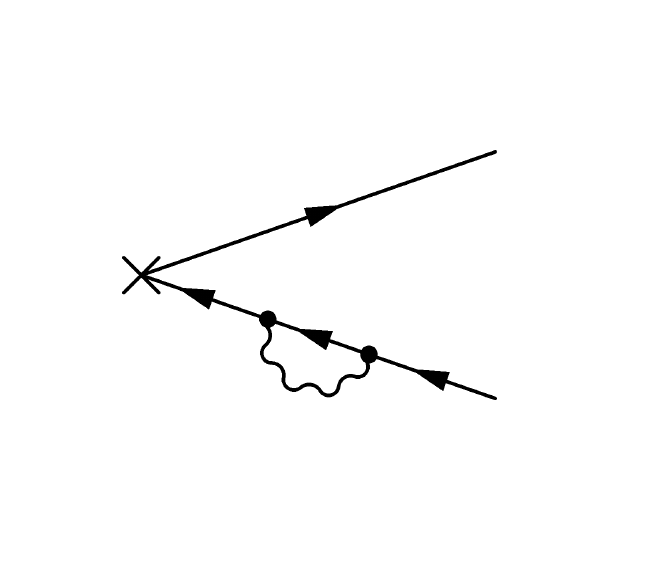} \\
\end{tabular}
\caption{The cross denotes the operator insertion. 
In the case of bare perturbation theory, $\delta Z_{\psib\psi} - \delta Z_\psi$
times the tree level diagram must cancel the loop divergences of the
other three diagrams.  In the case of renormalized perturbation theory,
the divergences will be cancelled by counterterms that are shown
in Fig.~\ref{psibpsictfig} below.
\label{psibpsidiag}}
\end{center}
\end{figure}

\subsubsection{Bare perturbation theory}
Our first
approach will be to use bare perturbation theory with dimensional regularization.
Then the finite Green function is obtained from the bare one according to
\beq
\langle 0 | T \psib \psi (x) \psi(y) \psib(z) | 0 \rangle
= Z_{\psib\psi} Z_\psi^{-1} \langle 0 | T \psib_0 \psi_0 (x) \psi_0(y) \psib_0(z) | 0 \rangle
\eeq
We will work to order $e^2$, and then writing $Z_\psi=1+\delta Z_\psi$ and $Z_{\psib\psi}
= 1 + \delta Z_{\psib\psi}$ this formula becomes
\beq
&& \langle 0 | T \psib \psi (x) \psi(y) \psib(z) | 0 \rangle
= \langle 0 | T \psib_0 \psi_0 (x) \psi_0(y) \psib_0(z) | 0 \rangle_{tree}
\ddd \qquad + \langle 0 | T \psib_0 \psi_0 (x) \psi_0(y) \psib_0(z) | 0 \rangle_{1-loop}
\ddd \qquad + (\delta Z_{\psib\psi} - \delta Z_\psi) \langle 0 | T \psib_0 \psi_0 (x) 
\psi_0(y) \psib_0(z) | 0 \rangle_{tree}
\label{olcgam}
\eeq
Thus the infinities in the 1-loop term must be subtracted off by corresponding
infinite values of $\delta Z_{\psib\psi}$ and $\delta Z_\psi$.  The tree diagram
is shown in the first diagram of Fig. \ref{psibpsidiag} and the 1-loop diagrams are shown in 
the other diagrams of Fig. \ref{psibpsidiag}.
On the other hand, $\delta Z_\psi$ is independently determined by the self energy
calculation
\beq
&& \langle 0 | T \psi(y) \psib(z) | 0 \rangle
= Z_\psi^{-1} \langle 0 | T \psi_0(y) \psib_0(z) | 0 \rangle
\ddd \qquad = \langle 0 | T \psi_0(y) \psib_0(z) | 0 \rangle_{tree} 
+ \langle 0 | T \psi_0(y) \psib_0(z) | 0 \rangle_{1-loop}
\ddd \qquad \qquad - \delta Z_{\psi} \langle 0 | T \psi_0(y) \psib_0(z) | 0 \rangle_{tree}
\label{bse}
\eeq
Comparing the two
calculations, it can be seen that the $-\delta Z_\psi$ in \myref{olcgam} cancels
one of the two self energy corrections show in Fig.~\ref{psibpsidiag}, but the other is not
cancelled, and so contributes to the value of $\delta Z_{\psib\psi}$.  This is
because at one loop \myref{olcgam} has twice as many self energy corrections
as \myref{bse}, but the same factor $Z_\psi^{-1}$.  The amputated version
of the upper right-hand diagram in Fig.~\ref{psibpsidiag}
will be denoted $\Gamma(q^2,p^2,q \cdot p)$, 
having passed to momentum space.  What we have found
is that
\beq
\delta Z_{\psib\psi} = -\Gamma(\mu) - \delta Z_\psi
\label{dZeq}
\eeq
where $\Gamma(q^2,p^2,q \cdot p) = \Gamma(-\mu^2,-\mu^2,-\mu^2) \equiv \Gamma(\mu)$ 
is the subtraction point.
What is left is to compute the two terms on the right-hand side.

Amputating the self energy diagram we obtain
\beq
-i \Sigma(\nott{p}) &=& (-ie)^2 \int \frac{d^dk}{(2\pi)^d} \frac{i \gamma_\mu (\nott{k}+\nott{p}) \gamma^\mu}{(k+p)^2 + i \e}
\frac{-i}{k^2+i\e} \nnn
&=& (d-2) e^2 \int \frac{d^dk}{(2\pi)^d} \frac{\nott{k}+\nott{p}}{[(k+p)^2 + i \e][k^2+i\e]} \nnn
&=& (d-2) i e^2 \nott{p} \int_0^1 dx ~ (1-x) (4\pi)^{-d/2} \Gamma(2 - \frac{d}{2}) \Delta^{\frac{d}{2}-2} \nnn
&=& \frac{ie^2}{8\pi^2} \nott{p} \int_0^1 dx ~ (1-x) ( \frac{2}{\e} + 1 - \ln \Delta - \gamma + \ln 4\pi )
\eeq
where $\Delta= -x(1-x)p^2$ and $d=4-\e$.  The subtraction condition arising from \myref{bse} is
\beq
\[ - \delta Z_\psi \frac{i}{\nott{p}} + \frac{i}{\nott{p}} (-i \Sigma(\nott{p})) \frac{i}{\nott{p}} \]_{\nott{p} = i\mu} = 0
\eeq
so that we find
\beq
\delta Z_\psi = - \frac{e^2}{8\pi^2} \int_0^1 dx ~ (1-x) ( \frac{2}{\e} + 1 - \ln [x(1-x)\mu^2] - \gamma + \ln 4\pi )
\eeq
The quantity that we will actually need is
\beq
\frac{d}{d\ln\mu} \delta Z_\psi = \frac{e^2}{8\pi^2}
\eeq

Next we consider the amputated diagram that was denoted above as $\Gamma(q^2,p^2,q \cdot p)$.  It is given
by 
\beq
\Gamma(q^2,p^2,q \cdot p) &=& (-ie)^2 \int \frac{d^d k}{(2\pi)^d} \gamma_\mu 
\frac{i ( \nott{k} + \nott{q} + \nott{p} ) }{ (k+q+p)^2 + i\e } 
\frac{i ( \nott{k} + \nott{p} ) }{ (k+p)^2 + i\e } \gamma^\mu 
\frac{-i}{k^2 + i\e}
\eeq
The numerator can be simplified to
\beq
\gamma_\mu (\nott{k}+\nott{q}+\nott{p}) (\nott{k}+\nott{p}) \gamma^\mu =
4 (k+q+p)\cdot (k+p) - \e (\nott{k}+\nott{q}+\nott{p}) (\nott{k}+\nott{p})
\eeq
and the $\ord{\e}$ term does not contribute to the singularity, but only to finite parts; therefore, we will not include it
in the subtraction and drop it henceforth, denoting the resulting integral with a prime, $\Gamma \to \Gamma'$.
It is now a simple matter to introduce Feynman parameters
and write the integral as
\beq
\Gamma'(q^2,p^2,q \cdot p) &=& 8e^2 \int_0^1 dx \int_0^{1-x} dy \int \frac{d^d {\tilde \ell}}{(2\pi)^d}
\frac{{\tilde \ell}^2 + {\tilde \Delta}}{({\tilde \ell}^2 + \Delta)^3}
\eeq
where $\ell = k + x q + (x+y) p$ is a shifted integration momentum, ${\tilde \ell}$ is this
momentum Wick rotated, 
\beq
\Delta &=& -x(1-x)q^2 -(x+y)(1-x-y)p^2 -2x(1-x-y)p\cdot q \nnn
{\tilde \Delta} &=& - x(1-x)q^2 - (1-x-y)^2 p^2 + (1-2x)(1-x-y)p\cdot q
\eeq
The term involving ${\tilde \Delta}$ is again finite, we drop it and denote the remaining integral
by a double-prime, wich evaluates to
\beq
\Gamma''(q^2,p^2,q \cdot p) = \frac{e^2}{2\pi^2} \int_0^1 dx \int_0^{1-x} dy \( \frac{2}{\e}
- \gamma - \half - \ln\Delta + \ln 4\pi \)
\eeq
Evaluating at the subtraction point $\Gamma''(q^2,p^2,q \cdot p) = \Gamma''(-\mu^2,-\mu^2,-\mu^2) \equiv \Gamma(\mu)$, 
we obtain
\beq
\frac{d}{d\ln\mu} \Gamma(\mu) = -\frac{e^2}{2\pi^2}
\eeq

Finally, taking into account \myref{dZeq},
\beq
\gamma_{\psib\psi} = -\frac{d}{d\ln\mu} \delta Z_{\psib\psi}
= \frac{d}{d\ln\mu} [ \Gamma(\mu) + \delta Z_\psi ] =  -\frac{e^2}{2\pi^2} + \frac{e^2}{8\pi^2} = -\frac{3e^2}{8\pi^2}
\eeq
Thus we see that the calcuation in bare pertubation theory agrees with \myref{unigam}
taking into account that the quadratic Casimir for a U(1) gauge theory with a fermion
of unit charge is $C_2(R)=1$.

\begin{figure}
\begin{center}
\begin{tabular}{ccc}
\includegraphics[width=2in]{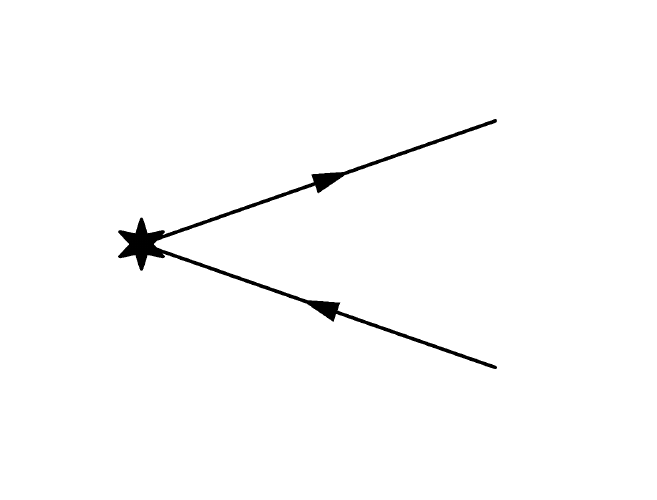} &
\includegraphics[width=2in]{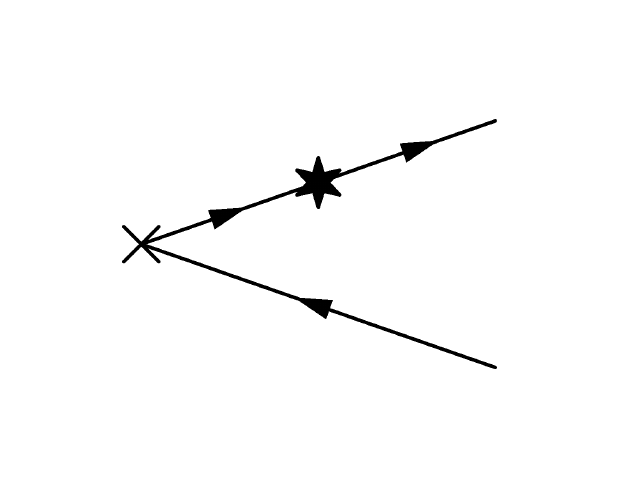} &
\includegraphics[width=2in]{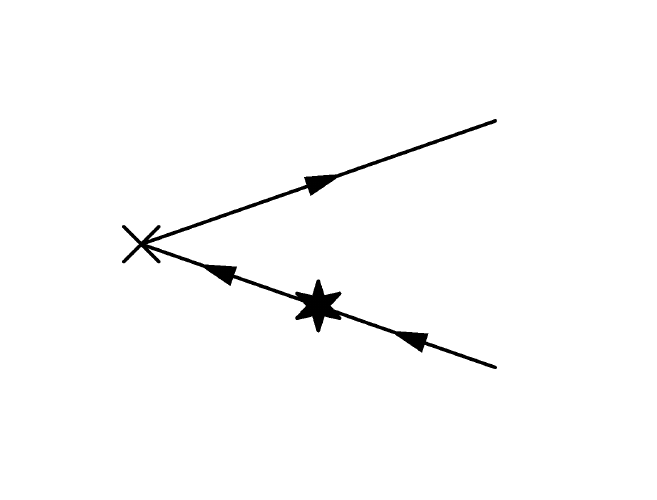}
\end{tabular}
\caption{Counterterms that cancel the loop divergences of Fig.~\ref{psibpsidiag} in
renormalized perturbation theory. Note that there are two self-energy
subtractions, in contrast to bare perturbation theory.  \label{psibpsictfig}}
\end{center}
\end{figure}

\subsubsection{Renormalized perturbation theory}
It is instructive to repeat this exercise from the point of view of renormalized
perturbation theory, since there it has an intimate connection with the Callan-Symanzik
equation.  It will be seen that we do not have any new integrals to compute---only their
interpretation is modified.  The additional counterterm diagrams are shown in
Fig.~\ref{psibpsictfig}.

We do not get $\gamma_{\psib \psi}$ directly from \myref{ganomi}, but
rather from the Callan-Symanzik equation
\beq
\( \mu \frac{\p}{\p \mu} + \beta(g) \frac{\p}{\p g} + 2 \gamma_{\psi} + \gamma_{\psib \psi} \)
G^{(2,1)}(p,q) = 0
\eeq
At the one loop level, this reduces to the following equation on the counterterms:
\beq
\mu \frac{\p}{\p \mu} \delta_{Z_{\psib \psi}} + 2 \mu \frac{\p}{\p \mu} \delta_{Z_2}
+ 2 \gamma_\psi + \gamma_{\psib \psi} = 0
\eeq
On the other hand, the Callan-Symanzik equation for the two point function
\beq
\( \mu \frac{\p}{\p \mu} + \beta(g) \frac{\p}{\p g} + 2 \gamma_{\psi}  \)
G^{(2)}(p) = 0
\eeq
yields
\beq
\mu \frac{\p}{\p \mu} \delta_{Z_2}
+ 2 \gamma_\psi = 0
\eeq
Using this to eliminate the anomalous dimension $\gamma_\psi$ of the elementary
field, we obtain
\beq
\gamma_{\psib\psi} = - \mu \frac{\p}{\p \mu} \( \delta_{Z_{\psib \psi}} + \delta_{Z_2} \)
\eeq

The counterterm $\delta_{Z_2}$ cancels both of the infinities from the external
leg corrections, so that in this formalism $\delta_{Z_{\psib \psi}}$ is purely
cancelling the divergence of the loop around the operator insertion.  Thus
we recover what we found in bare perturbation theory:  the anomalous
mass dimension comes from that loop plus one self-energy loop.  We obtain
an identical result.

\subsubsection{Mass independent scheme}
Obviously we could repeat all of the calculations with a nonzero fermion mass $m$.  However,
what we will now show is that the $Z_{\psib\psi}$ and $Z_\psi$ used in the bare perturbation
theory in the massless case work just as well in the massive case, as far as rendering the
correlation function finite goes.  Thus one arrives at a mass independent scheme for
renormalizing the bare correlation function in the massive case.  This has generalizations
to other calculations of renormalization constants which we shall comment on at the
end of this discussion.

Consider the two 1-loop diagrams that had infinities, but now for the massive theory.
For instance, we now have
\beq
&& \Gamma(q^2,p^2,q \cdot p) =
\ddd
(-ie)^2 \int \frac{d^d k}{(2\pi)^d} \gamma_\mu 
\frac{i ( \nott{k} + \nott{q} + \nott{p} + m ) }{ (k+q+p)^2 - m^2 + i\e } 
\frac{i ( \nott{k} + \nott{p} + m) }{ (k+p)^2 - m^2 + i\e } \gamma^\mu 
\frac{-i}{k^2 + i\e}
\eeq
The numerator ``simplification'' now becomes
\beq
&& \gamma_\mu (\nott{k}+\nott{q}+\nott{p} + m) (\nott{k}+\nott{p} + m) \gamma^\mu =
\ddd 4 (k+q+p)\cdot (k+p) - \e (\nott{k}+\nott{q}+\nott{p}) (\nott{k}+\nott{p})
\ddd -(2-\e) m (2\nott{k} + 2\nott{p} + \nott{q}) + d m^2
\eeq
where again $d=4-\e$.  It is now clear that all of the terms proportional to $m$ come with one lower power
of the loop momentum and are therefore finite, since the leading integral is log divergent.  Similarly
the term involving $m^2$ is finite.  Therefore none of these mass dependent terms are divergent, and
they do not require a subtraction in order to make the correlation function finite.

Next we expand the denominator in powers of $m^2$.  For instance,
\beq
\frac{1}{ (k+q+p)^2 - m^2 + i\e } = \frac{1}{ (k+q+p)^2 + i\e } \[ 1 + {\cal O} \( \frac{m^2}{(k+q+p)^2 + i\e} \) \]
\eeq
Obviously all of the corrections are suppress by additional powers of $1/k^2$ and therefore finite,
again because the leading integral is log divergent.  Thus we reach a similar conclusion that mass dependent
terms do not require subtraction.

The same line of argument applies to the self energy calculation because it is also log divergent.
We find therefore that expanding in powers of the fermion mass $m$, the mass dependent terms are
finite, and the only subtractions that we need are in the mass independent part.  As a result
we can use the $Z_{\psib\psi}$ and $Z_\psi$ determined from the massless theory to render the
massive theory finite.  Hence we reach the conclusion that a mass independent
renormalization scheme is possible.

The key ingredient here was that the leading divergence is logarithmic, a fact that extends to
higher orders by power counting.
If we were renormalizing some other operator, and we found loop diagrams with quadratic divergences, 
then expanding in $m$, the $m^2$ term would almost certainly contain a logarithmic divergence.  In that
case the Z factor for the operator would have to contain a mass dependent term in order to renormalize
its correlation functions.  Thus it is possible to have a situation where a mass independent
scheme cannot be achieved if what one is interested in is the operator renormalization constant
in the massive theory.  Of course this does not change the fact that it is always possible to
write the anomalous dimensions in a mass-independent way, if they are defined through the
Callan-Symanzik equation with $m \p / \p m$ essentially counting mass insertions into massless
Green's functions, multiplied by $\gamma_m$.  For instance see Eq.~\myref{withmass} below.

As a specific example, the dimension five
operator $\Ocal_5^a = {\bar q} \gamma_5 \tau^a D_\mu D^\mu  q$ has
this difficulty, where $q=(u,d)$ are quark fields.  
In the lattice application in which this arises (explained shortly),
the operator is multiplied by the lattice spacing $a$.  The leading divergence 
associated with $a \Ocal_5^a$ is then
linear and the term proportional to $m$ will be logarithmically divergent.
Hence it will require a mass-dependent subtraction.  In fact this operator is very interesting
to study because it is the one generated from an axial flavor transformation of the
action with Wilson quarks, and hence appears in the axial Ward identity
\beq
\langle \p_\mu J_{5\mu}^a (x) \Ocal(y) \rangle
= \langle a \Ocal_5^a(x) \Ocal(y) \rangle
\eeq
The mass dependent
renormalization of this operator in fact is responsible for the multiplicative renormalization
proportional to the bare mass, whereas the linear divergence is responsible for the
additive renormalization:  $m_r = Z_m (m_0 - m_c)$ with $m_c \sim 1/a$ and $Z_m \sim \ln a$.

\subsection{Dilatation Ward identities}
The anomalous dimension is associated with scale transformations; thus it is useful
to review the associated Ward identities.
The scale transformation ($x \to \lambda x$) of a field is given by
\beq
\Phi(x) \to \Phi'(x) = \lambda^{-\Delta} \Phi(x/\lambda)
\eeq
The infinitesmal form is obtained by taking $\lambda = 1 + \e$ with $\e \ll 1$:
\beq
\delta \Phi(x) = -\e (\Delta + x^\mu \p_\mu) \Phi(x)
\eeq
The global form of the Ward identity is obtained from the path integral by
taking $\e =$ const., using the identity (change of variables of integration invariance)
\beq
\int [d\Phi] e^{-S[\Phi]} \Phi_1(x_1) \cdots \Phi_n(x_n)
= \int [d\Phi'] e^{-S[\Phi']} \Phi'_1(x_1) \cdots \Phi'_n(x_n)
\eeq
with $\Phi'_i = \Phi_i + \delta \Phi_i$ and expanding in $\e$ to linear order,
since $\e$ is infinitesmal.  Assuming that the
action and measure are invariant under the scale transformation
\beq
[d\Phi'] = [d\Phi], \quad S[\Phi'] = S[\Phi]
\eeq
this yields
\beq
&& \int [d\Phi] e^{-S[\Phi]} \Phi_1(x_1) \cdots \Phi_n(x_n) 
= \int [d\Phi] e^{-S[\Phi]} \Phi_1(x_1) \cdots \Phi_n(x_n)
\ddd \qquad - \e \int [d\Phi] e^{-S[\Phi]} \sum_i \Phi_1(x_1) \cdots \Phi_{i-1}(x_{i-1}) ( \Delta_i + x_i^\mu \p_\mu^{(i)} )
\Phi_i(x_i) 
\ddd \qquad \times \Phi_{i+1}(x_{i+1}) \cdots \Phi_n(x_n)
\eeq
Thus the quantity that $\e$ multiplies must vanish, which is the Ward identity
\beq
\sum_i \langle 0 | T \{ \Phi_1(x_1) \cdots \Phi_{i-1}(x_{i-1}) ( \Delta_i + x_i^\mu \p_\mu^{(i)} )
\Phi_i(x_i) \Phi_{i+1}(x_{i+1}) \cdots \Phi_n(x_n) \} | 0 \rangle = 0
\label{gdid}
\eeq

Similarly, there is a local form of the Ward identity involving the dilatation
current $D^\mu$, which is obtained by taking $\e = \e(x)$, i.e., a local
transformation.  This is not an invariance of the action, but rather
\beq
S[\Phi'] = S[\Phi] + \int d^4 x ~ \p_\mu \e(x) D^\mu(x)
\eeq
I.e., it would be invariant if $\p_\mu \e(x)=0$.
Repeating steps as above, we find that
\beq
0 &=& \int [d\Phi] e^{-S[\Phi]} \int d^4 y ~ \p_\mu \e(y) D^\mu(y) \Phi_1(x_1) \cdots \Phi_n(x_n)
\ddd + \int [d\Phi] e^{-S[\Phi]} \sum_i \Phi_1(x_1) \cdots \Phi_{i-1}(x_{i-1}) 
\e(x_i) ( \Delta_i + x_i^\mu \p_\mu^{(i)} ) \Phi_i(x_i) 
\ddd \qquad \times \Phi_{i+1}(x_{i+1}) \cdots \Phi_n(x_n)
\eeq
Taking the functional derivative $\delta/\delta \e(x)$ of this equation, using
$\delta \e(y) / \delta \e(x) = \delta(x-y)$, one obtains
\beq
0 &=& - \frac{\p}{\p x^\mu} \( \int [d\Phi] e^{-S[\Phi]} D^\mu(x) \Phi_1(x_1) \cdots \Phi_n(x_n) \)
\ddd + \int [d\Phi] e^{-S[\Phi]} \sum_i \Phi_1(x_1) \cdots \Phi_{i-1}(x_{i-1}) 
\delta(x-x_i) ( \Delta_i + x_i^\mu \p_\mu^{(i)} ) \Phi_i(x_i)
\ddd \qquad \times \Phi_{i+1}(x_{i+1}) \cdots \Phi_n(x_n)
\eeq
We then find that
\beq
&& \frac{\p}{\p x^\mu} \langle 0 | T \{  D^\mu(x) \Phi_1(x_1) \cdots \Phi_n(x_n) \} | 0 \rangle
\ddd \qquad = \sum_i \langle 0 | T \{ \Phi_1(x_1) \cdots \Phi_{i-1}(x_{i-1}) 
\delta(x-x_i) ( \Delta_i + x_i^\mu \p_\mu^{(i)} ) \Phi_i(x_i) 
\ddd \qquad \times \Phi_{i+1}(x_{i+1}) \cdots \Phi_n(x_n) \} | 0 \rangle
\label{ldid}
\eeq
Integration of this relation over $x$ yields the global identity \myref{gdid}.

This local Ward identity can also be obtained purely from operator manipulations and current algebra.
For the sake of simplicity, let us assume $x_1^0 > x_2^0 > \cdots > x_n^0$
so that the operator product $\Phi_1(x_1) \cdots \Phi_n(x_n)$ is already
time-ordered.  Then
\beq
&& T \{  D^\mu(x) \Phi_1(x_1) \cdots \Phi_n(x_n) \} 
= \theta(x^0 - x_1^0) D^\mu(x) \Phi_1(x_1) \cdots \Phi_n(x_n)
\ddd \qquad + \theta(x_1^0 - x^0) \theta(x^0 - x_2^0) \Phi_1(x_1) D^\mu(x) \Phi_2(x_2) \cdots \Phi_n(x_n)
\ddd \qquad + \cdots + \theta(x_{n-1}^0 - x^0) \theta(x^0 - x_n^0) \Phi_1(x_1) \cdots \Phi_{n-1}(x_{n-1})
D^\mu(x) \Phi_n(x_n) 
\ddd \qquad + \theta(x_n^0 - x^0) \Phi_1(x_1) \cdots \Phi_n(x_n) D^\mu(x)
\eeq
We apply $\p / \p x^\mu$ to this and impose conservation of the current, $\p_\mu D^\mu = 0$.
We also make use of the identities
\beq
\frac{\p}{\p x^0} \theta(x^0 - y^0) = -\frac{\p}{\p x^0} \theta(y^0 - x^0) = \delta(x^0 - y^0)
\eeq
This results in
\beq
&& \frac{\p}{\p x^\mu} \langle 0 | T \{  D^\mu(x) \Phi_1(x_1) \cdots \Phi_n(x_n) \} | 0 \rangle
\ddd = \langle 0 | T \{ \delta(x^0 - x_1^0) [ D^0(x) , \Phi_1(x_1) ] \Phi_2(x_2) \cdots \Phi_n(x_n)
\ddd \qquad + \Phi_1(x_1) \delta(x^0 - x_2^0) [ D^0(x) , \Phi_2(x_2) ] \Phi_3(x_3) \cdots \Phi_n(x_n)
\ddd \qquad + \cdots + \Phi_1(x_1) \cdots \Phi_{n-1}(x_{n-1}) \delta(x^0 - x_n^0) [ D^0(x), \Phi_n(x_n) ]
\} | 0 \rangle
\label{widmanip}
\eeq
The integral over space of $D^0$ gives the dilatation charge, and the equal time commutator
of the charge with a given field gives its transformation with respect to dilatations.
Thus we have the current algebra relation\footnote{Minkowski time and Euclidean time
are related by $t_M = -i t_E$.  It follows that since we demand that the Minkowski
space current conservation $\p_\mu^{M} D_M^\mu=0$ becomes $\p_\mu^{E} D_\mu^E=0$, the temporal
component of the dilatation current is related in the two formulations by
$D_M^0 = -i D_0^E$.  Defining the dilation charge as $Q_{M,E} = \int d^3 x ~ D_{M,E}^0$,
the Minkowski space relation $i [ Q_M, \Phi(x) ] = \delta \Phi(x)$ becomes
$[Q_E, \Phi(x)] = \delta \Phi(x)$. It is for this reason there is no factor
of $i$ in Eq.~\myref{curralg}.}
\beq
\delta(x^0-y^0) [ D^0(x) , \Phi(y) ] = \delta^{(4)}(x-y) ( \Delta + y^\mu \p_\mu^{(y)} ) \Phi(y)
\label{curralg}
\eeq
where we made explicit that the delta function on the r.h.s.~is four-dimensional.
Substituting this into \myref{widmanip} yields \myref{ldid}.

In a theory with a quantum scale anomaly, the path to the renormalized relation
from the bare theory depends on the regulator.  For instance, in dimensional
regularization, the trace of the energy momentum tensor is no longer zero in
$d$ dimension, and a term proportional to $d-4$ appears.  In the lattice theory,
the energy momentum tensor is no longer a conserved current due to the
explicit violation of translation invariance, and hence it mixes with other
operators.  This leads to a renormalized energy momentum tensor that is
not traceless.  
  Whatever the regulator, additional terms on the right-hand
side of the above equation are generated in the renormalized theory.  The
specific form of these depend on theory.  For instance, in a gauge theory
the additional term is proportional to
\beq
\langle 0 | F_{\mu\nu}^2 (y) \Phi_1(x_1) \cdots \Phi_n(x_n) | 0 \rangle
\eeq
In $\phi^4$ theory the additional term involves the $\phi^4$ operator \cite{Wilson:1970wp}.
It is interesting that the operator dimensions $\Delta$ still appear in
a Ward identity even when scale invariance is violated in the quantum
theory, so that in principle if one knew the Green functions one could obtain
the anomalous dimensions.

\section{Hyperscaling}
\label{sec-hyperscaling}
\subsection{Basic RG arguments}
The hyperscaling relation is simply derived from the renormalization group equations.  It has
recently been elucidated in \cite{DelDebbio:2010ze}, though the result is quite a bit older.
In the vicinity of a RG fixed point, a RG transformation modifies the parameters according to
\beq
\mu = \lambda \mu', \quad g = \lambda^{-y_g} g', \quad {\hat m} = \lambda^{-y_m} {\hat m}'
\label{coupling-rescale}
\eeq
As a reminder, $\hat m$ is the dimensionless mass, defined relative to the UV
cutoff; for instance on the lattice, $\hat m = m a$.
In our preliminary discussion, we will ignore the coupling $g$, since it is associated
with an irrelevant operator and would be zero\footnote{In this discussion,
the coupling $g$ is measured relative to the critical coupling at zero mass:
i.e., what is really being discussed is $g-g_c$.  Obviously this can
be dealt with with a simple redefinition $g \to g-g_c$.} if we only consider relevant
perturbations around the fixed point.  However, we will return to nonzero $g$
when we consider scaling violations below.
Taking into account the anomalous dimension of the correlation function
$C(t;\hat m,\mu) = \langle \Ocal_H(t) \Ocal_H^\dagger(0) \rangle$ of the
zero momentum operator 
\beq
\Ocal_H(t) = \lim_{L \to \infty} \frac{1}{L^3} \int d^3 x ~ \Ocal_H(t,{\bf x})
\eeq
under the RG,
\beq
C(t;\hat m,\mu) \approx \lambda^{-2 \gamma_H} C(t;{\hat m}',\mu')
\eeq
Again it must be emphasized that this simple scaling law is a property that
only holds in the neighborhood of a fixed point, and in fact since the theory
is always studied away from the fixed point, the behavior is asymptotic,
hence the symbol ``$\approx$'' as opposed to ``$=$.''  In general there are
``scaling violations'' present that should ultimately be taken into account
in any realistic study.
On dimensional grounds, since $\mu' t$ is the basic dimensionless quantity
we can form out of the dimensionful parameters $\mu'$ and $t$, and
the dimensions of the correlation function is $2d_H$,
\beq
C(t;{\hat m}',\mu') = (\mu')^{2d_H} F(\mu' t, {\hat m}')
= \lambda^{-2d_H} \mu^{2d_H} F(\mu (\lambda^{-1} t), {\hat m}')
= \lambda^{-2d_H} C(\lambda^{-1} t; {\hat m}', \mu)
\label{rescale-arg}
\eeq
Now we choose $\lambda$ such that ${\hat m}'=1$, which translates into $\lambda^{-1} = {\hat m}^{1/y_m}$.
It follows that
\beq
C(t;{\hat m},\mu) = {\hat m}^{2 \Delta_H / y_m} C({\hat m}^{1/y_m} t; 1, \mu)
\label{corrmhone}
\eeq
where $\Delta_H = d_H + \gamma_H$.
Since $C(t) \sim e^{-M_H t}$ is the time dependence we see that $M_H t \sim {\hat m}^{1/y_m} t$ and
then on dimensional grounds
\beq
M_H = \kappa_H \mu {\hat m}^{1/y_m}
\label{hyperscaling}
\eeq
where $\kappa_H$ is a dimensionless constant that is independent of ${\hat m}$.
We see that the scaling with ${\hat m}$ is independent of which physical state $H$ we examine.
This is the hyperscaling result:  all physical masses should have the same power law behavior.
In fact, not only should the leading exponential display this behavior, but so should the
subleading exponentials.  This implies that it is not only the ground state which has
hyperscaling, but also excited states, multiparticle states, etc.  Any energy eigenstate will
scale like ${\hat m}^{1/y_m}$.
Note that one prediction of the hyperscaling \myref{hyperscaling} is that ratios of masses
of the composite states will be constant as a function of the mass ${\hat m}$.
This constancy has been checked in a number of lattice gauge theories that
may or may not be inside the conformal window, as will be described in more detail in
this section below.
Another consequence is that there may not be a clear separation of states in the
spectrum.  Chiral symmetry is not spontaneously broken, since we cannot have a dynamical
scale in a scale invariant theory, so there are no Nambu-Goldstone bosons.  However,
we do not even have a guaranteed hierarchy of scales between the flavor nonsinglet pseudoscalar states
and the rest of the spectrum, so a chiral effective lagrangian at nonzero explicit
chiral symmetry breaking (${\hat m} \not= 0$) may not make any sense.\footnote{In fact there is at least
one example where the scalar flavor singlet state seems to be lighter
than the ``pions'' \cite{Aoki:2013hla}.}  Of course there will
still be an effective low energy theory, but it may contain many fields in addition to
ones representing the flavor nonsinglet pseudoscalar states.  It is certainly a rich
topic to confront with lattice simulation data.

An important further result derived in \cite{DelDebbio:2010ze} is that the chiral
condensate scales as
\beq
\vev{\psib \psi} \sim {\hat m}^{\eta_{\psib \psi}}, \quad \eta_{\psib \psi} = \frac{3 - \gamma_*}{1 + \gamma_*}
\label{condscaling}
\eeq
Another result that they derive is that the gluon condensate scales as
\beq
\vev{G^2} \sim {\hat m}^{\eta_{G^2}}, \quad \eta_{G^2} = \frac{4}{1+\gamma_*}
\eeq
The probelm with this is that there is a short distance singularity
$\vev{G^2} \sim a^{-4}$ that will overwhelm the effect.  Using
Wilson flow would avoid this problem \cite{Luscher:2010iy}.

The condensate scaling \myref{condscaling} can be seen from the considerations
above.  Choosing $H = \psib \psi$ in \myref{corrmhone}, and taking into account
that as $t \to \infty$ the correlation function is saturated by the vacuum,
\beq
C(t;{\hat m},\mu) = {\hat m}^{2 \Delta_{\psib\psi} / y_m} C({\hat m}^{1/y_m} t; 1, \mu)
\sim {\hat m}^{2 \eta_{\psib \psi}}
\eeq
which implies that
\beq
\eta_{\psib\psi} = \frac{ \Delta_{\psib\psi} }{ y_m } = \frac{3 - \gamma_*}{1 + \gamma_*}
\eeq

In fact, for any operator $\Ocal$ with quantum numbers of the vacuum, one has
the possibility of a nonzero vacuum expectation value $\langle 0 | \Ocal(x) | 0 \rangle$.
On the other hand, in the limit of large $|x|$, the correlation function is
saturated by the vacuum,
\beq
\langle 0 | \Ocal(x) \Ocal(0) | 0 \rangle \to \langle 0 | \Ocal(0) | 0 \rangle^2
\sim {\hat m}^{2\eta_\Ocal} \sim {\hat m}^{2\Delta_\Ocal / y_m}
\eeq
where the last behavior follows from the RG arguments that are given above.  So it
is a general property of ``condensates.''

The relation between the mass scaling of the condensate and the Dirac operator eigenvalue
density $\rho(\lambda)$ is also straightforward.  
In the Euclidean spacetime in the continuum, the eigenvectors of the massless Dirac operator $D$ have
eigenvalues which are purely imaginary, which we write as 
\beq
D \varphi_{\lambda,n} = i \lambda \varphi_{\lambda,n}
\eeq
where $n$ labels different eigenvectors with the same eigenvalue $i \lambda$;
i.e., we admit the possibility of degeneracies, which will be important shortly.
Then adding a mass,
\beq
(D + m) \varphi_{\lambda,n} = (i \lambda + m) \varphi_{\lambda,n}
\eeq
Taking into account the completeness relation 
\beq
\sum_{\lambda,n} \varphi_{\lambda,n}(x) \varphi_{\lambda,n}^\dagger(y) = \delta(x-y) {\bf 1}
\eeq
where $x,y$ are spacetime coordinates, it is easy to see that
\beq
(D + m)^{-1}(x,y) = \sum_{\lambda,n} \frac{ \varphi_{\lambda,n}(x) \varphi_{\lambda,n}^\dagger(y) }
{ i \lambda + m }
\eeq
On the other hand, the condensate is given by
\beq
\vev{ \psib \psi } = -\frac{1}{V_4} \sum_x \langle 0 |  \psi_\alpha(x) \psib_\alpha(x) | 0 \rangle
= -\frac{1}{V_4} \sum_x \tr (D+m)^{-1}(x,x)
\eeq
where $\alpha$ is the spinor index (it could also include flavor if $N_f > 1$), $V_4$ is
the spacetime volume, and a minus sign enters because of
the anticommutation of the fermions.  Using the representation in terms of eigenvectors and eigenvalues,
we therefore find
\beq
\vev{\psib \psi} = -\frac{1}{V_4} \sum_x \sum_{\lambda,n} 
\tr \frac{ \varphi_{\lambda,n}(x) \varphi_{\lambda,n}^\dagger(x) }
{ i \lambda + m }
\eeq
The eigenvectors are normalized to unity so that
\beq
\sum_x \tr \varphi_{\lambda,n}(x) \varphi_{\lambda,n}^\dagger(x)
= \sum_x \varphi_{\lambda,n}^\dagger(x) \varphi_{\lambda,n}(x) = 1
\eeq
Thus the condensate reduces to
\beq
\vev{\psib \psi} = - \frac{1}{V_4} \sum_{\lambda,n} \frac{1}{ i \lambda + m }
\eeq
For a given value of $\lambda$, the sum over $n$ divided by $V_4$ gives the
density of eigenvalues $\rho(\lambda)$.  The formula becomes
\beq
\vev{\psib \psi} = - \sum_\lambda \frac{\rho(\lambda)}{i \lambda + m}
\eeq
Transitioning to infinite spacetime volume, the eigenvalues
become continuous and we have
\beq
\vev{\psib \psi} = \int_{-\infty}^{\infty} d \lambda ~ \frac{\rho(\lambda)}{i \lambda + m}
\eeq
Since at this point we are considering the continuum massless Dirac operator $D$, 
\beq
\{ D , \gamma_5 \} = 0
\eeq
It then follows that
\beq
D \varphi_{\lambda,n} = -\gamma_5 D \gamma_5 \varphi_{\lambda,n} = i\lambda \varphi_{\lambda,n}
\eeq
from which it follows that
\beq
D ( \gamma_5 \varphi_{\lambda,n} ) = - i\lambda ( \gamma_5 \varphi_{\lambda,n} )
\eeq
Thus, for every nonzero $\lambda$ and for every $n$, there is a corresponding
eigenvector $\varphi_{-\lambda,n} = \gamma_5 \varphi_{\lambda,n}$ with
eigenvalue $-i \lambda$.  (The zeromodes are chiral, $\gamma_5 \varphi_{0,n} =
\pm \varphi_{0,n}$, so that $\gamma_5$ multiplication does not generate
a linearly independent vector for $\lambda=0$.)  It follows that
the eigenvalue density is an even function, $\rho(-\lambda)=\rho(\lambda)$, so
we can further simplify as follows
\beq
\vev{\psib\psi} = -\int_0^\infty d \lambda ~ \( \frac{\rho(\lambda)}{i \lambda + m}
+ \frac{\rho(\lambda)}{-i \lambda + m} \)
= -2 m \int_0^\infty d \lambda ~ \frac{\rho(\lambda)}{\lambda^2 + m^2}
\eeq
The condensate has UV divergences that must be subtracted off to obtain
a renormalized quantity, in addition to a multiplicative renormalization.
This can be viewed as a mixing with the identity operator ${\bf 1}$,
since $\psib\psi$ has the quantum numbers of the vacuum.  Thus from
an operator renormalization perspective
\beq
(\psib\psi)_r = Z_S (\psib\psi)_0 + Z_{{\bf 1}} {\bf 1}
\eeq
However, on dimensional grounds $Z_{{\bf 1}}$ must have mass dimension 3.
In the continuum with a dimensionless regulator such as dimensional regularization,
this mixing with a lower dimensional operator would vanish in the massless
theory.  Thus we know that in the case of dimensional regularization
\beq
Z_{{\bf 1}} = c_3 m^3
\eeq
where $c_3 \sim 1/\e$.  One might wonder whether it is also possible to
have a term $c_1 \mu^2 m$ with $c_1 \sim 1/\e$ but this would require
a loop divergence of the form 
\beq
\frac{1}{\e} p^2 m
\eeq
in order to yield the factor of $\mu^2$ at the subtraction point $p^2 = -\mu^2$.
However, the condensate only gets a contribution from $p=0$ and so such
a subtraction does not occur for this operator.  On the other hand with
a dimensionful regulator such as the lattice, where $\Lambda_{UV} = 1/a$,
or Pauli-Villars where $\Lambda_{UV}$ is the Pauli-Villars mass, one
can also have the term linear in mass, so that
\beq
Z_{{\bf 1}} = c_1 \Lambda_{UV}^2 m + c_3 m^3
\eeq
where $c_1$ and $c_3$ are functions of $\ln(\Lambda_{UV}/m)$.
Thus for instance with the lattice regulator the renormalized
condensate is related to the density by
\beq
\vev{\psib\psi} = -2m \int_0^\mu d\lambda ~ \frac{\rho(\lambda)}{\lambda^2+m^2}
-2m^5 \int_\mu^\infty \frac{d\lambda}{\lambda^4} \frac{\rho(\lambda)}{\lambda^2+m^2}
+ c_1 a^{-2} m + c_3 m^3
\label{fullcond}
\eeq
where $\rho(\lambda) = Z_S \rho_0(\lambda)$ is the rescaled eigenvalue density.
All terms except the first term on the right-hand side of \myref{fullcond}
are analytic in the mass, and we write them as $A(m)$.  Then by a simple
change of variables
\beq
\vev{\psib\psi} = -2 \int_0^{\mu/m} dx ~ \frac{\rho(mx)}{x^2+1} + A(m)
\eeq
If for small values of $\lambda$ we have a power law $\rho \sim \lambda^\alpha$
then in the $m \to 0$ limit
\beq
\vev{\psib\psi} \sim m^\alpha \int_0^\infty dx ~ \frac{x^\alpha}{x^2+1} + A(0)
\eeq
Since $\vev{\psib\psi} \sim m^{\eta_{\psib\psi}}$, we see that in the $\lambda \to 0$ limit
\beq
\rho(\lambda) \sim \lambda^{\eta_{\psib\psi}}
\eeq

It is also interesting to consider the amplitude in the correlation function, and not just
the exponent.  This is particularly true in the case of the appeal to volume reduction
(translation group orbifold equivalence).  Here, a single site lattice can be used to
study the infinite volume theory, \`a la Eguchi-Kawai.  However, the observables in the
infinite volume theory are those that are invariants of the lattice translation group,
for instance $\sum_{x,y} \langle 0 | \Ocal(x) \Ocal^\dagger(y) | 0 \rangle$, where $x,y$
are four-dimensional site labels.  This obviously corresponds to a susceptibility and
will (i) include all of the excited states (because the early $t$ values are included in the integral)
and (ii) does not give access to the exponential
decay directly.
Let's look at a susceptibility in some detail, vis-\`a-vis the usual correlation functions.
Then we have:
\beq
C(t) = \sum_{{\bf x}} \langle 0 | \Ocal(t,{\bf x}) \Ocal^\dagger(0,{\bf 0}) | 0 \rangle
\eeq
Then the susceptibility is just
\beq
\chi = \sum_t C(t)
\eeq
We will first carry all of this out in infinite volume and then worry about finite boxes.
As usual, the correlation function can be written in terms of states:
\beq
C(t) = \sum_i A_i e^{-t m_{H_i} }, \quad A_i = | \langle 0 | \Ocal(0,{\bf 0}) | H_i \rangle |^2
\label{corrgen}
\eeq
where $H_i$ represents a ``hadronic'' state, which may actually be a multiparticle state in many cases.
So first since we are in infinite volume we only care about hyperscaling --- i.e., the dependence on
the current (PCAC) quark mass $m$.
We say from the RHS of \myref{corrmhone} that 
\beq
C({\hat m}^{1/y_m} t; 1, \mu) = \sum_i {\hat A}_i(\mu) e^{-\kappa_{H_i} \mu {\hat m}^{1/y_m} t}
\eeq
so that ${\hat A}_i(\mu)$ are independent of ${\hat m}$.  Since that is true then in \myref{corrgen}
\beq
A_i = {\hat m}^{2 \Delta_H / y_m} {\hat A}_i(\mu)
\eeq
and we could extract the exponent ratio $\Delta_H/y_m$ from examining how the amplitude depends on
${\hat m}$.
But now notice that the index $i$ has fallen off of $H$ in $\Delta_H$.  We are assuming that
the operator $\Ocal$ has dimension $\Delta_H$.  But in general things are not so simple, because
the operator will mix with other operators under RG flow.  What one would need to do is to find
an ``eigenoperator'' basis,\footnote{This terminology has also been used by Fisher
with a similar meaning \cite{Fisher15}.} where the operator has a well-defined dimension and does not mix.
It may be that the ground state corresponds to just this sort of operator, in which case in
the large $t$ behavior of the correlation function, we only ``see'' one operator dimension, even
though we do not necessarily have in hand an eigenoperator.
Where in this whole line of reasoning does the IRFP make its appearance?  It is in
the idea that operators, masses, couplings, scale with $\lambda$ to some power when $\mu \to \mu' = \lambda^{-1} \mu$.
In the limit of no mixing, we find for the susceptibility in the zero temperature limit
\beq
\lim_{\beta \to \infty} \int_0^\beta dt ~ C(t;{\hat m},\mu) 
= \sum_i {\hat A}_i(\mu) \frac{1}{\kappa_{H_i} \mu} {\hat m}^{(2 \Delta_H - 1)/y_m}
\eeq
So by extracting the scaling of the susceptibility w.r.t.~${\hat m}$ we can obtain the
critical exponent $(2 \Delta_H - 1)/y_m$.  This should be tested explicitly in the
model of $SU(N)$ gauge theory with two Dirac fermions in the adjoint representation,
since many strands of evidence point to this having an IRFP, and many other studies
indicate that the center symmetry will remain unbroken in the $b = 1/g^2 N \to \infty$
limit, so that the volume reduction should hold.

\subsection{Including mixing}
We now modify the above arguments to take into account mixing.
To do this, we expand our ``hadronic'' operator $\Ocal_H$ in terms
of the eigenoperators that have definite anomalous dimension and do
not mix:
\beq
\Ocal_H = \sum_i c_i \Phi_i
\label{expphi}
\eeq
We will denote the anomalous dimension of $\Phi_i$ as $\gamma_i$.
Then the correlation function also has an expansion
\beq
C_H(t) &=& \lim_{L\to \infty} \frac{1}{L^3} \sum_{{\bf x}} \langle 0 | \Ocal_H(t,{\bf x}) \Ocal_H^\dagger(0,{\bf 0}) | 0 \rangle
\ddd = \sum_{ij} c_i c_j^* C_{ij}(t)
\eeq
where
\beq
C_{ij}(t) = \lim_{L\to \infty} \frac{1}{L^3} \sum_{{\bf x}} \langle 0 | \Phi_i(t,{\bf x}) \Phi_j^\dagger(0,{\bf 0}) | 0 \rangle
\eeq
The scaling relation under an RG transformation is now replaced by
\beq
C_{ij}(t;{\hat m},\mu) \approx \lambda^{-(\gamma_i + \gamma_j)} C_{ij}(t;{\hat m}',\mu')
\eeq
Thus it is no longer the case that $C_H(t)$ is simply rescaled, since
\beq
C_H(t;{\hat m},\mu) \approx \sum_{ij} c_i c_j^* \lambda^{-(\gamma_i + \gamma_j)} C_{ij}(t;{\hat m}',\mu')
\eeq
However, if we proceed as in the previous subsection without mixing, we find
\beq
C_H(t;{\hat m},\mu) \approx \sum_{ij} c_i c_j^* {\hat m}^{(\Delta_i+\Delta_j)/y_m} C_{ij}({\hat m}^{1/y_m} t, 1 , \mu)
\eeq
Hence due to the time dependence on the r.h.s., hyperscaling $M_H \sim {\hat m}^{1/y_m} \mu$ will still hold.

So far we have only considered ultra-local operators.  However, on the lattice
this is too restrictive since in the continuum limit operators that contain
fields that are separated by distances of order the lattice spacing also
become local operators.  Thus in \myref{expphi} for the case of $P^+ = {\bar u} d$
one should also include operators such as ${\bar u}(x) P \exp ( i \int_x^y dz^\mu A_\mu(z) ) d(y)$
with $|| x - y || \sim \ord{a}$ and the ``P'' in front of the exponential denoting
path ordering.  Of course on the lattice, the integral is replaced
with a product of link fields $U_\mu(z)$ that make a path from $x$ to $y$.
Exactly this sort of operator is used in what is known as smearing,
where the quark fields ${\bar u}$ and $d$ are replaced by smeared
versions that are translated over the lattice in a gauge covariant
way using the link fields.  This is often used to suppress the
excited states in the correlation function.  Alternatively, it
is used to build a large basis of operators so that a variational
analysis can be carried out in order to access the excited states
in a numerically precise way.  Here, in the context of hyperscaling,
how might smearing be used?  One answer is of course that the mass $M_H$
can be more reliably extracted, as usual, by eliminating excited state
contamination.  There may be other ways to exploit smearing and
$\ord{a}$-improved operators for the purposes of determining critical
exponents; this should certainly be explored.

\subsection{Irrelevant operators}
Now we repeat the steps leading to \myref{hyperscaling} keeping the irrelevant
coupling $g$ that appears in \myref{coupling-rescale}, which is zero at the
IR fixed point.\footnote{In the gauge theory, we have redefined the coupling
so that it will vanish at the fixed point, $g \to g - g_*$, where $g_*$ is the
fixed point coupling in the original formulation.} 
We now have
\beq
C(t;\hat m,g,\mu) \approx \lambda^{-2 \gamma_H} C(t;{\hat m}',g',\mu')
\eeq
This follows from the Callan-Symanzik equation
\beq
\[ \mu \frac{\p}{\p \mu} + \beta(g) \frac{\p}{\p g} + 2 \gamma_H + \gamma_m {\hat m} \frac{\p}{\p {\hat m}} \] C(t;\hat m, g, \mu) =0
\label{withmass}
\eeq
where
\beq
C(t;\hat m,g,\mu) = Z_H^2 \langle 0 | \Ocal_{H,0}(t) \Ocal_{H,0}^\dagger(0) | 0 \rangle
\eeq
Here, $\Ocal_{H,0}$ is the bare ``hadronic'' operator.
Near the fixed point $\gamma_H \approx$ const.~so that
\beq
\frac{Z_H(\mu')}{Z_H(\mu)} \approx \lambda^{\gamma_H}
\eeq
according to \myref{ganomi}.
Repeating the steps in \myref{rescale-arg}, we find that
\beq
C(t;{\hat m}',g',\mu') = \lambda^{-2d_H} C(\lambda^{-1} t; {\hat m}', g', \mu)
\eeq
Again taking $\lambda^{-1} = {\hat m}^{1/y_m}$, we find
\beq
C(t;{\hat m},g,\mu) = {\hat m}^{2 \Delta_H / y_m} C({\hat m}^{1/y_m} t; 1, {\hat m}^{-y_g/y_m} g, \mu)
\eeq
where we remind the reader that $\Delta_H = d_H + \gamma_H$.
It follows that the mass extracted from this correlator will be given by
\beq
M_H = \mu {\hat m}^{1/y_m} f({\hat m}^{-y_g/y_m} g)
\label{nzg}
\eeq
At the fixed point $g=0$ we simply recover $\kappa_H = f(0)$ to achieve
agreement with \myref{hyperscaling}.  Away from the fixed point the
dependence is more complicated than $M_H \sim {\hat m}^{1/y_m}$.  Modifications
like this are beginning to be considered in lattice studies \cite{Cheng:2013xha}, since one always
deals with the irrelevant gauge coupling which flows very slowly near the
fixed point, and hence the precise fixed point behavior ($g=0$) is not
seen without some degree of scaling violation contamination.  We will discuss this issue
in the section on finite size scaling, Sec.~\ref{fss}, below, where the study \cite{Cheng:2013xha}
provides an example of taking into account corrections to scaling due to
an irrelevant coupling that is flowing very slowly to zero.

\subsection{General property of the unstable fixed point}
The point $m=0$ is an unstable fixed point.  If we allow $m$ to be slightly nonzero,
the theory will flow away to a large mass $m'$ after RG blocking.  Following an
argument found in \cite{Cardy96} for the Ising model, we can see that the masses
$M_H$ of states in the spectrum must go to zero as we approach the fixed
point, consistent with the prediction of hyperscaling.  The point is that associated
with $M_H$ is a correlation length, $\xi_H = 1/M_H$.  Under an RG blocking
with blocking factor $s$, i.e., $a' = sa$, the correlation length will
be shortened according to $\xi_H(m') = \xi_H(m)/s$.  So let us suppose
a reference mass $m_0$ with $\xi_H(m_0) = \xi_0 = \ord{1}$.  If we start
the flow with $m < m_0$, it will take $n(m)$ steps for $\xi_H$ to reach
the value $\xi_0$, i.e., $\xi_H(m) = s^{n(m)} \xi_0$.  But as we take
$m$ closer to the fixed point $m=0$, more and more steps are required.
Eventually, $n(m) \to \infty$ as $m \to 0$.  Thus $\xi_H(m) \to \infty$.
It follows that $M_H(m) \to 0$ as $m \to 0$.

\subsection{Hyperscaling determination of the anomalous mass dimension from the lattice}
This approach requires that the infinite spatial volume extrapolation be
made.  There are various levels of sophistocation in performing this
extrapolation.  An example is found for ``minimal walking technicolor''
in \cite{DelDebbio:2013hha}.  There it is found that it is necessary to
push to $m_\pi L \gappeq 10$ in order to avoid finite volume corrections.
In fact this is typical of the ``beyond the standard model'' applications:
the quantity $m_\pi L$ must be quite a bit larger than in lattice QCD.
In \cite{DelDebbio:2013hha} the anomalous dimension obtained from hyperscaling
is found to be consistent with the mode number analysis to be discussed below.
Hyperscaling was applied to the twelve flavor model in \cite{Fodor:2011tu} and \cite{Aoki:2012eq}
reaching conclusions that were at odds with each other.  The first article
found a ``very low level of confidence in the conformal scenario'' where
all masses and decay constants would obey the hyperscaling relation,
driven by the anomalous mass dimension.  In \cite{Aoki:2012eq} the authors
report that the twelve flavor theory is consistent with the ``conformal
hypothesis'' and find $\gamma \sim 0.4$ to $0.5$.  In a subsequent article \cite{Aoki:2013hla}
this group measured the sigma mass; the ratio $m_\sigma/m_\pi$ should be
a constant with respect to fermion mass $m$; however, in Table 1 it is seen
that this quantity is not quite constant, at least for the smaller volumes.
If the conformal hypothesis is correct then this could possibly be blamed
on finite volume corrections.  It would be interesting to see a finite-size
scaling analysis of these results.
This conclusion is aided by the results of \cite{Aoki:2014oha} by the same group,
which shows a straight-line relationship between $m_\sigma^2$ and $m_\pi^2$
in the data.  Strangely, $m_\sigma^2$ extrapolates to a negative value
in the chiral limit.
  They have recently reported further studies
supportive of hyperscaling \cite{Aoki:2015gea}.  A recent study \cite{Lombardo:2014pda}
of the twelve flavor theory by another group includes, among other things, an
extrapolation to infinite volume and an extraction of $\gamma$ from
the hyperscaling relation.  All hadron masses yield a consistent value
of $\gamma\sim 0.2$.

\section{Deconstruction}
In the ``deconstruction'' approach \cite{Stephanov:2007ry}, the composite operator $\Ocal$ is
resolved in terms of elementary excitations.
We first note that the correlation function has a spectral decomposition
\beq
\int d^4 x ~ e^{i p \cdot x} \langle 0 | \Ocal(x) \Ocal^\dagger(0) | 0 \rangle
= \int \frac{dM^2}{2\pi} \frac{ i \rho_\Ocal(M^2) }{ p^2 - M^2 + i \e}
\label{KL1}
\eeq
Since the scaling dimension of the left-hand side is $2 \Delta_\Ocal - 4$,
we find that near a conformal fixed point the spectral density must satisfy
\beq
\rho_\Ocal(M^2) = A_\Ocal \cdot (M^2)^{\Delta_\Ocal - 2}
\label{rhoscaling}
\eeq
where $A_\Ocal$ is a constant.  On the other hand, if the scale symmetry is slightly
broken (for instance with a small fermion mass $m$ or finite size $L$), 
then the operator will have a discrete spectrum [not including the continuum
of momentum states already accounted for in \myref{KL1}]
\beq
\rho_\Ocal(M^2) = 2\pi \sum_n \delta(M^2 - M_n^2) | \langle 0 | \Ocal(0) | \varphi_n \rangle |^2
\label{rhodecomp}
\eeq
Of course \myref{KL1} and \myref{rhodecomp} taken together just comprise the standard
K\"all\'en-Lehmann spectral representation of the two-point function in terms of
energy eigenstates $| \varphi_n \rangle$ with zero momentum (cf.~for instance
Section 7.1 of \cite{Peskin95}).

This description can be related to a decomposition of the operator in terms of 
``eigenoperators'' that create the elementary excitations,\footnote{Note that
there is a connection here to the variational approach that is utilized in
lattice gauge theory.}
\beq
\Ocal = \sum_n f_n \varphi_n
\eeq
The elementary excitations thus have corresponding decay constants
\beq
| \langle 0 | \Ocal(0) | \varphi_n \rangle |^2 \equiv f_n^2
\eeq

In the deconstruction analysis of the decomposition,
one imagines that the spectrum has to do with the Kaluza-Klein
spectrum of an extra dimension.
The spacing between the spectrum is controlled by a parameter $\delta$
with mass dimension $[\delta]=1$.  One possible choice, taken in \cite{Stephanov:2007ry},
(we will consider others below---including one more natural to the
extra dimensional perspective) is
\beq
M_n^2 = n \delta^2
\label{stephsplit}
\eeq
If we study the limit $\delta \to 0$, where the sum in \myref{rhodecomp} becomes
an integral, and match it to the scaling behavior \myref{rhoscaling}, then we find
for the decay constants
\beq
f_n^2 = \frac{A_\Ocal}{2\pi} \delta^2 (M_n^2)^{\Delta - 2}
\eeq

The presence of the scale symmetry breaking mass $m$ allows for
a nonvanishing vacuum expectation value (condensate) of the operator
$\Ocal$.  Hence in the effective Lagrangian there is a linear term
in $\Ocal$, with a nonzero coefficient that depends on $m$ in
such a way that it vanishes in the $m \to 0$ limit.
Because of the anomalous dimension, the $m^{1/y_m}$ has mass
dimension 1 under scale transformations, and hence to construct
a dimensionless term in the action, the effective Lagrangian must take the form
\beq
\Lcal_\text{eff} &=& \ldots  -c m^{(4-\Delta)/y_m} \Ocal - \half \sum_n M_n^2 \varphi_n^2 \nnn
&=& \ldots  - c m^{(4-\Delta)/y_m} \sum_n f_n \varphi_n - \half \sum_n M_n^2 \varphi_n^2
\eeq
where the ``$\ldots$'' represents terms that are not important to
the arguments that we make here, and $c$ is a dimensionless constant.
From this the equations of motion give
\beq
\vev{\varphi_n} = - \frac{c m^{(4-\Delta)/y_m} f_n}{M_n^2}
\eeq
It then follows that
\beq
\vev{\Ocal} = \sum_n f_n \vev{\varphi_n} = - c m^{(4-\Delta)/y_m} \sum_n \frac{f_n^2}{M_n^2}
\eeq
In the small $\delta$ limit (which should be valid as the
conformal symmetry is restored as $m \to 0$) the sum can be evaluated as
\beq
 \sum_n \frac{f_n^2}{M_n^2} = \sum_n \frac{ (n\delta^2)^{\Delta-2} }{n}
= \sum_n \delta^2 (n \delta^2)^{\Delta-3} \approx \int ds ~ s^{\Delta-3}
\eeq
Imposing IR and UV cutoffs on the integral, we have
\beq
\vev{\Ocal} \approx -cm^{(4-\Delta)/y_m} \int_{\Lambda_{IR}^2}^{\Lambda_{UV}^2} ds ~ s^{\Delta-3}
= - \frac{ cm^{(4-\Delta)/y_m} }{ \Delta-2 } ( (\Lambda_{UV}^2)^{\Delta-2} - (\Lambda_{IR}^2)^{\Delta-2} )
\eeq
The IR cutoff should be of order $M_{n=1}$ and we expect 
$M_{n=1}$ to obey hyperscaling,
$M_{n=1} \sim m^{1/y_m}$.  In terms of the IR dependence, we thus have
\beq
\vev{\Ocal} |_{IR} \sim m^{(4-\Delta)/y_m} (m^{2/y_m})^{\Delta-2} = m^{\Delta/y_m}
\label{cscl}
\eeq
One shortcoming of this approach is that it does not explain why the UV contribution
would be analytic, since it seems to scale as $m^{(4-\Delta)/y_m}$.
However we know that this is not the case since the critical behavior that
leads to nontrivial anomalous dimensions is due to IR singularities; hence
we have to imagine that a more rigorous approach would smooth out the behavior
in the UV.

Now let us consider, instead of \myref{stephsplit}, the more typical case
of Kaluza-Klein spectrum,
\beq
M_n^2 = n^2 \delta^2
\eeq
On dimensional grounds,
\beq
f_n^2 = c \delta^{2 + 2 \nu} (M_n^2)^{\Delta_\Ocal - 2 - \nu}
\eeq
with $c$ and $\nu$ dimensionless constants that need to be
determined.  Again we study the limit $\delta \to 0$, where the sum in \myref{rhodecomp} becomes
an integral, and match it to the scaling behavior \myref{rhoscaling}.  Then
\beq
&& \rho_\Ocal(M^2) = 2 \pi \int dn ~ \delta(M^2 - M_n^2) f_n^2
= 2 \pi c \delta^{1 + 2 \nu} \int dM_n ~ \delta(M^2 - M_n^2) (M_n^2)^{\Delta_\Ocal - 2 - \nu}
\ddd = \pi c \delta^{1 + 2 \nu} (M^2)^{\Delta_\Ocal - 2 - \nu - \half}
\eeq
Thus we conclude that $\nu = -\half$ and $c = A_\Ocal / \pi$ so that
\beq
f_n^2 = \frac{A_\Ocal}{\pi} \delta (M_n^2)^{\Delta_\Ocal - \frac{3}{2}}
\eeq
Then we return to the calculation of the condensate, as above, and find
\beq
&& \vev{\Ocal} = \sum_n f_n \vev{\varphi_n} = -m^{(4-\Delta)/y_m} \sum_n \frac{f_n^2}{M_n^2}
= -m^{(4-\Delta)/y_m} \int dn ~ \delta \frac{A_\Ocal}{\pi} (n^2 \delta^2)^{\Delta_\Ocal - \frac{3}{2} - 1}
\ddd = -m^{(4-\Delta)/y_m} \frac{A_\Ocal}{\pi} \int d \sqrt{s} ~ s^{\Delta_\Ocal - \frac{5}{2}}
= -m^{(4-\Delta)/y_m} \frac{A_\Ocal}{2\pi} \int ds ~ s^{\Delta_\Ocal - 3}
\eeq
so we reach the same conclusion as before, Eq.~\myref{cscl}.  Thus the prediction of
this ``deconstruction'' is robust in terms of the scaling dimension of the condensate, and
the choice of elementary excitation spectrum is unimportant.

\section{Finite size scaling}
\label{fss}
\subsection{General arguments}
Since 1971 it has been known that there is a scaling theory for the
smoothing of thermodynamic singularities by finite size effects \cite{Fisher71},
with further foundational work in the following year \cite{Fisher:1972zza}.
For a lattice system of size $L$, there are essentially three scales:
the lattice spacing $a$, the correlation length $\xi$, and the system
size $L$.  Thermodynamic quantities can depend on the dimensionless
ratios $\xi/a$ and $L/a$.  The finite size scaling hypothesis states
that close to the critical point, the quantities become independent
of the microscopic scale $a$, and thus can only depend on the
ratio $\xi/L$.  For instance, the correlation length
in system size $L$, which we denote $\xi_L$, will have the form
\beq
\xi_L = \xi \phi(\xi/L) = L \frac{\xi}{L} \phi(\xi/L) \equiv L F(\xi/L)
\eeq
where $\phi(0)=1$ in order to recover the infinite volume result.
Given the hyperscaling relation $\xi \sim m^{-1/y_m}$, we therefore
obtain
\beq
\xi_L = L f(m^{1/y_m} L)
\label{sfg}
\eeq
as a finite size scaling relation.  Thus we can extract $y_m=1+ \gamma_*$
by fitting a scaling curve through the data.  Note that it has been
assumed that we are close to the critical point, which in our application
means that we have a theory with an IRFP and $m$ is sufficiently small.

An alternative derivation, which is based on the renormalization group, follows
what was done above in Section \ref{sec-hyperscaling} for hyperscaling.  Under
an RG transformation,
\beq
C_H(t;{\hat m},L,\mu) = \lambda^{-2\gamma_H} C_H(t;{\hat m}',L,\mu')
\eeq
where Eqs.~\myref{coupling-rescale} still hold.
Paying attention to the dimensionless ratios that we can form and writing
all dimensionful dependence in terms of $\mu'$, on the right-hand side
\beq
&& C_H(t;{\hat m}',L,\mu') = (\mu')^{2d_H} F(t \mu', {\hat m}', L \mu')
= \lambda^{-2d_H} \mu^{2d_H} F((\lambda^{-1} t) \mu, {\hat m}', (\lambda^{-1} L) \mu)
\ddd = \lambda^{-2d_H} C_H(\lambda^{-1} t, {\hat m}', \lambda^{-1} L, \mu)
\eeq
Thus we find that
\beq
C_H(t;{\hat m},L,\mu) = \lambda^{-2\Delta_H} C_H(\lambda^{-1} t, \lambda^{y_m} {\hat m}, 
\lambda^{-1} L, \mu)
\eeq
Choosing again ${\hat m}' = \lambda^{y_m} {\hat m} = 1$, so that $\lambda = {\hat m}^{-1/y_m}$,
we see that the right-hand side of this equality becomes
\beq
\lambda^{-2\Delta_H} C_H(\lambda^{-1} t, \lambda^{y_m} {\hat m}, 
\lambda^{-1} L, \mu) = \mu^{2d_H} {\hat m}^{2\Delta_H/y_m} F(\mu t {\hat m}^{1/y_m}, 1 ,
{\hat m}^{1/y_m} L \mu)
\eeq
But this is also supposed to equal
\beq
C_H(t;{\hat m},L,\mu) = A e^{-m_H t} + B e^{-m_{H^*} t} + \cdots
\eeq
In order for this to match for all $t$, we must have
\beq
m_H = \kappa_H \mu {\hat m}^{1/y_m} \phi_H({\hat m}^{1/y_m} L \mu), \quad
m_{H^*} = \kappa_{H^*} \mu {\hat m}^{1/y_m} \phi_{H^*}({\hat m}^{1/y_m} L \mu), \quad
\cdots
\label{fssmass}
\eeq
Thus the mass is given by the hyperscaling relation up to the scaling
functions $\phi({\hat m}^{1/y_m} L \mu) = f(x)$ where $x = m^{1/y_m} L$
is the scaling variable (in units of $\mu$).  Again we see that excited
states also enjoy a finite size scaling.

Once again, we consider the limits of the finite size scaling equations
\myref{fssmass}.  In order to obtain hyperscaling in the
thermodynamic limit $L \to \infty$, we must have $\phi(\infty)=1$.
On the other hand, we know that as ${\hat m} \to 0$ we should
get vanishing masses $m_H, m_{H^*}, \ldots$, hence $\phi(0) = $ finite.

\subsection{Determining $y_m$ from fits to interpolating functions}
Here the method is the one used in \cite{DeGrand:2009hu,DeGrand:2011cu,Giedt:2012rj},
originating in \cite{Bhatta}.  It attempts to optimize $y_m$ such
that all the data falls on a scaling curve.
The simulation is carried out on a lattice of size $L^3 \times T$,
where $T = \zeta L$ is the size in the temporal direction, and
typically the aspect ratio $\zeta=2$.  (Note that we must beware of the possibility
that the scaling function depends on $\zeta$.)
For each $L$ we have a data set $p$ corresponding to a collection of different
values of the PCAC mass $m$.  We use this
to obtain a fit $f_p$.  The types of fit functions
that we considered in \cite{Giedt:2012rj} will be described below.  We then
use this fit function on the other values of $L$, which
we label as $L_j$. 

We minimize the following function with respect to $y_m$.
\beq
P(y_m) = \frac{1}{N_{\text{over}}} \sum_p \sum_{j \not= p} \sum_{i,\text{over}}
\( \frac{\xi_L(m_{i,j})}{L_j} - f_p(L_j^{y_m} m_{i,j}) \)^2
\label{tomin}
\eeq
Here $i$ labels the different partially conserved axial current (PCAC) mass values for a given $L_j$.
The effect of this is to find a $y_m$ such that $f_p$ for the
other values $L_j, m_{i,j}$ is as close as possible to the
curve obtained from fitting $L_p, m_{i,p}$.  This is summed
over all possibilities $p$.  Also, ``over'' indicates that
only $i$ are used such that the scaling variable $x_{i,j} = m_{i,j} L_j^{y_m}$ falls within
the range of values of $x_{i,p} = m_{i,p} L_p^{y_m}$, so that the comparison
is to an interpolation of the $x_{i,p}$ data,
rather than an extrapolation.  Unweighted fits were
used so that the approximation to the scaling curve
would pass through data at small $x=m L^{y_m}$,
where absolute (statistical) errors are largest.  (Using
a weighted fit reduced our conclusion for $\gamma$ by 4\%.)

\begin{table}
\begin{center}
\begin{tabular}{|c|c|} \hline\hline
Type & $f(x)$ \\ \hline \hline
Quadratic & $c_0+c_1 x+c_2 x^2$ \\ \hline
Log quadratic & $c_0 + c_1 \ln x + c_2 (\ln x)^2$ \\ \hline
Piece-wise log-linear & Straight lines connecting data \\ \hline\hline
\end{tabular}
\caption{Interpolating functions that we use to fit data for
a fixed $L_p$.  In the last case, the straight lines
interpolating between data are on a semi-log plot. \label{tabfit}
}
\end{center}
\end{table}

For the fitting function we considered the possibilities
listed in Table \ref{tabfit}.  In the case of the quadratic 
we follow one of the methods of \cite{DeGrand:2009hu,DeGrand:2011cu}.
The log quadratic fit was motivated by the behavior of
the data when $\xi_L/L$ is plotted versus $\ln(m L^{y_m})$,
which is close to a parabola.  The piece-wise log-linear
form was used as a third choice that trivially passes
through the data, giving a reasonable interpolation.

We have used four observables:  the ``pion'' mass $m_\pi$,
the ``rho'' mass $m_\rho$, the ``$a_1$'' mass $m_{a_1}$,
and the ``pion'' decay constant $f_\pi$.  These are all
obtained from standard correlation functions using
point sources and sinks.  We fit the correlation functions
with a single exponential, allowing the first time $t_{\text{first}}$ in the fit
to be large enough for the excited state contributions
to be small.  This is determined by looking at the mass
of the meson as a function of $t_{\text{first}}$ and
extracting the value on the plateau.  Five values of
bare masses $m_0 a = -1.0, -1.1, -1.165, -1.175, -1.18$
on lattices of size $L/a = 10, 12, 16, 20, 24$ were
simulated, all at $\beta=2.25$.  These are the same configurations as were
generated in \cite{Giedt:2011kz}, and the values of
the PCAC mass and details on the simulations are given there.  
Also note that the size of the temporal direction is $T=2L$.

\begin{figure}
\begin{center}
\includegraphics[width=4in]{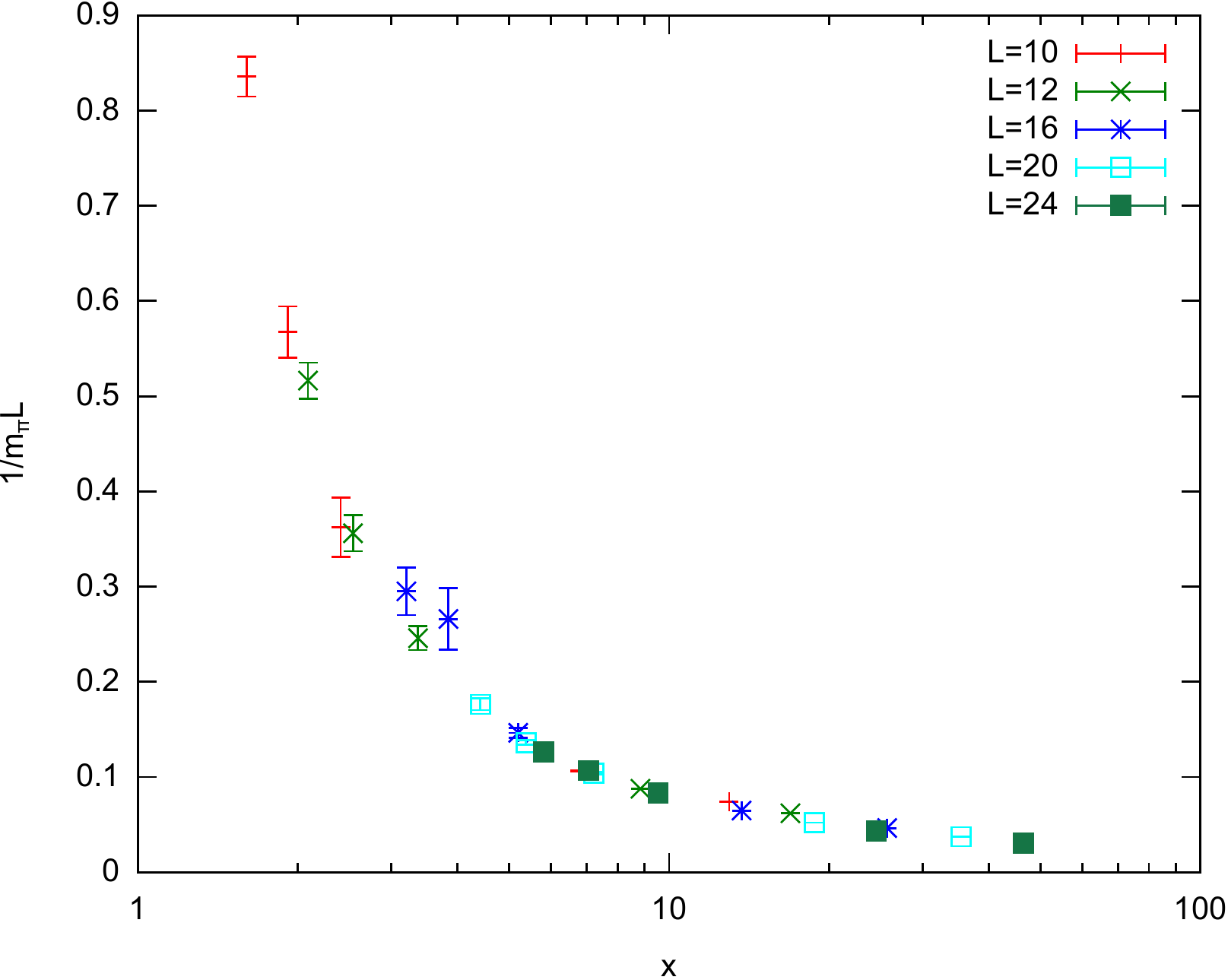}
\caption{Collapse of $\pi$ data for $y_m=1.46$.  Here and
in the other figure, $x=m L^{y_m}$.
\label{fig1} }
\end{center}
\end{figure}

\begin{figure}
\begin{center}
\includegraphics[width=4in]{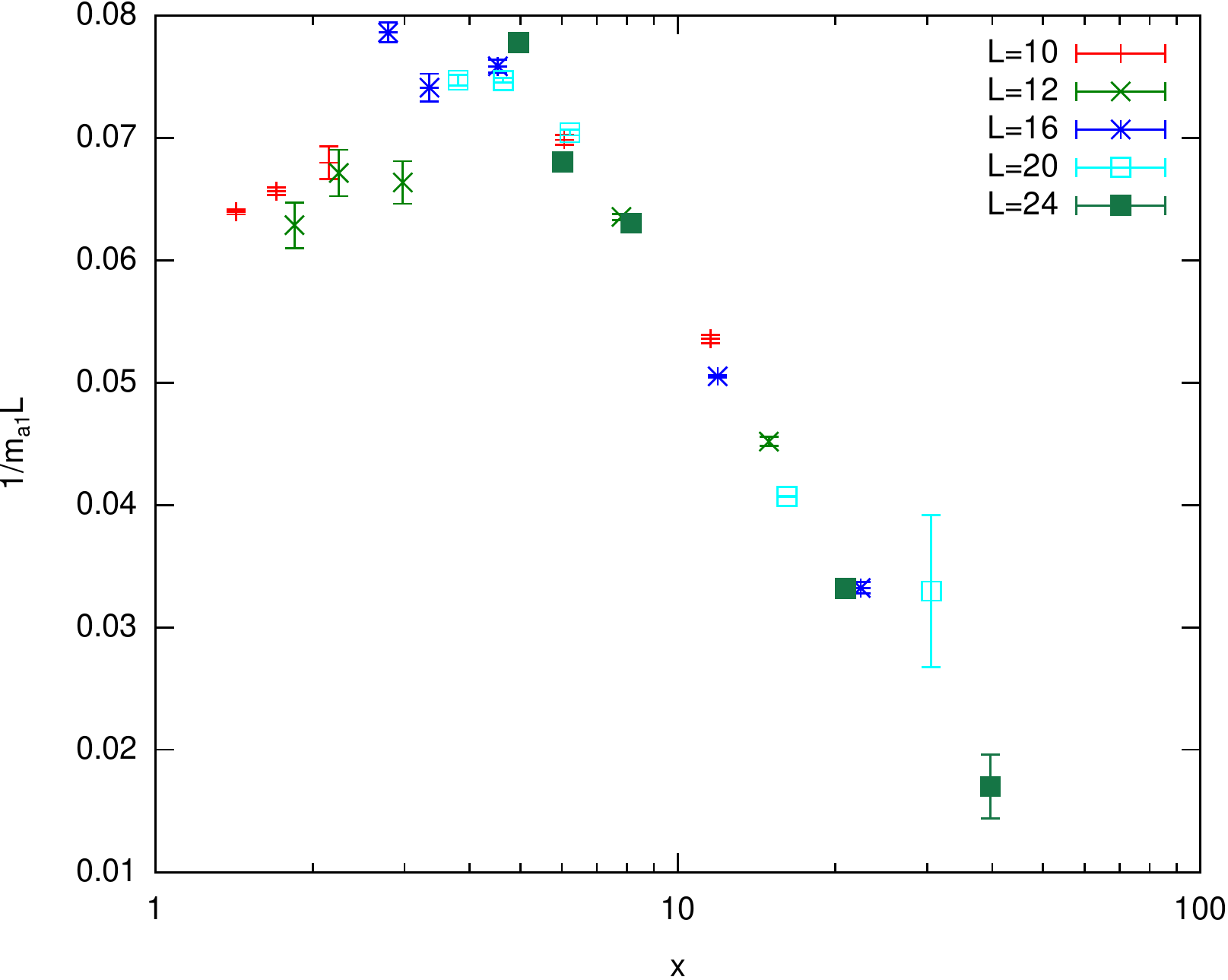}
\caption{The $a_1$ data for $y_m=1.41$.
\label{fig3} }
\end{center}
\end{figure}

Using these results, and performing the minimization
of \myref{tomin} described above, we obtain values for $y_m$.
In the case of $m_{a_1}$ and $f_\pi$, the quantity
$\xi_L/L$ is small, and scaling violations [cf.~Eq.~\myref{scavio}] can compete
with the scaling function for small lattices.  For this
reason we exclude the small lattices $L/a=10,12$ for these
channels.  The results for $y_m$ are summarized in
Table \ref{tabus}.  It can be seen that each of the
channels, and each of the fitting methods are consistent
with each other within errors.  
The approximate collapse of data in the pion channel is shown in Fig.~\ref{fig1};
the rho looks quite similar.  In Fig.~\ref{fig3} we show the
scatter that occurs for the $a_1$; for $f_\pi$ the spread
in data is similar. In both cases it is the small $L$
observables that are pulling away from the curve. 
We interpret this as being due to scaling violations,
though a thorough study extracting $\omega$ would be required to demonstrate this.
Another interpretation is that the theory does not have an IRFP,
and so the FSS fails for some channels.  It is also possible that
we are seeing the effect of $\beta=2.25$ not being close
enough to the fixed point coupling (also scaling violations).  We view the
collapse seen in Fig.~\ref{fig1}
as favoring our scaling violation interpretation.

\begin{table}
\begin{center}
\begin{tabular}{|c|c|c|c|} \hline \hline
Observable & Quadratic & Log Quad  &		PWL  \\ \hline \hline
$m_\pi$	  &	1.67(93) & 1.26(54)	&  1.51(33)  \\ \hline
$m_\rho$	& 1.67(88) & 1.37(39) &	 1.56(31)  \\ \hline
$m_{a_1}$	&	1.40(52) & 1.42(27) &  1.41(22)  \\ \hline
$f_\pi$   &	1.65(22) & 1.49(54) &	 1.60(29)  \\ \hline \hline
\end{tabular}
\caption{The scaling exponent $y_m=1+\gamma$ for the various observables and methods of
interpolation.  In parentheses, jackknife error is shown, obtained from eliminating
one $m_{i,j}$ in all possible ways, in the minimization of \myref{tomin}.  
Because we use a large number of configurations, $\ord{10^3}$,
statistical error is negligible by comparison.  Weighted averages
and standard deviations are shown in the last column.
\label{tabus} }
\end{center}
\end{table}

We average the twelve values of $y_m$ for the four channels and
three fitting methods, weighted by the jackknife errors, to obtain
$\overline{\gamma}=0.50$.  The standard deviation of
the twelve fits is 0.13.  However, the smallest jackknife
error from single elimination of data is 0.22.  Treated
as separate systematic errors, we obtain
\beq
\gamma = 0.50 \pm 0.26
\eeq
In Table \ref{taball}, we compare to other
results using variety of methods.  We are
in agreement with all but the FSS studies \cite{DelDebbio:2010hu,DelDebbio:2010hx},
though only 1.4$\sigma$ different from their upper limit.

\begin{table}
\begin{center}
\begin{tabular}{|c|c|} \hline \hline
Method & $\gamma$ \\ \hline \hline
SF \cite{Bursa:2009we} & $0.05 < \gamma < 0.56$ \\ \hline
SF \cite{DeGrand:2011qd} & $0.31 \pm 0.06$ \\ \hline
Perturbative 4-loop \cite{Pica:2010xq} & $0.500$ \\ \hline
Schwinger-Dyson \cite{Ryttov:2010iz} & $0.653$ \\ \hline
All-orders hypothesis \cite{Pica:2010mt} & $0.46$ \\ \hline
MCRG \cite{Catterall:2011zf} & $-0.6 < \gamma < 0.6$ \\ \hline
FSS \cite{DelDebbio:2010hu} & $0.05 < \gamma < 0.20$ \\ \hline
FSS \cite{DelDebbio:2010hx} & $0.22 \pm 0.06$ \\ \hline
Mode number \cite{Patella:2012da} & $0.371 \pm 0.020$ \\ \hline
FSS (here) & $0.50 \pm 0.26$ \\ \hline \hline
\end{tabular}
\caption{Summary of all MWTC results for the anomalous mass dimension.
SF is Schr\"odinger functional and MCRG is Monte Carlo renormalization group.
The perturbative result $\gamma=0.500$ was also given in the later, corrected version
of \cite{Ryttov:2010iz}, and relied on invariants calculated in \cite{Mojaza:2010cm}.
\label{taball} }
\end{center}
\end{table}

\subsection{Scaling of eigenvalues}
In \cite{DeGrand:2009hu} DeGrand has fit the low-lying
eigenvalues of the Dirac operator as a function of
the lattice size $L$, based on theoretical developments
in the older work \cite{Akemann:1997wi}.  The functional
form is given by
\beq
\vev{\lambda_i} \sim L^{-p}, \quad p = \frac{d}{1+\alpha}, 
\quad \alpha = \frac{d}{y_m}-1 =\frac{3-\gamma}{1+\gamma}
\label{pform}
\eeq
where we remind the reader that in our conventions $y_m=1+\gamma$
and $d=4$.
In DeGrand's calculation he simulates tree-level clover
improved sextet fermions and measures the eight lowest eigenvalues of
the overlap Dirac operator on these configurations.  As can
be seen on the log-log plot of Fig.~8 of \cite{DeGrand:2009hu},
a power law behavior is, broadly speaking, observed.  Careful
fitting is performed and it is found that the fit to a power law
behavior is superior if only the four lowest eigenvalues are
included.  The answers for $\gamma$ have rather small error
by this method, of order 3\%.  Concentrating on the three
largest volumes and the four smallest eigenvalues, $p = 1.59 \pm 0.05$.
After a bit of algebra using \myref{pform} above we find
that $p=y_m$ and hence $\gamma = 0.59 \pm 0.05$,
which is fairly consistent with what is found for the sextet model
using other methods, such as the Schr\"odinger functional
to be discussed below.

\subsection{Other studies using finite size scaling}
In \cite{DeGrand:2009hu} the FSS approach was found to yield $\gamma \approx 0.5$ for
the SU(3) theory with $N_f=2$ sextet fermions.  This is consistent with the Schr\"odinger
functional results discussed in the next section.  This paper also makes the important
point that the value of $\gamma$ obtained should be independent of the bare coupling
used in the simulation.\footnote{Of course the bare action must be such that the
theory is in the basin of attraction of the IRFP.}

In \cite{Athenodorou:2014eua}, the authors
use FSS to find $\gamma \sim 0.9$ for $N_f=1$ SU(2) adjoint.
This large value of $\gamma$ is encouraging phenomenologically,
and is similar to what was found for $N_f=10$ SU(3) fundamental \cite{Appelquist:2012nz}.
It seems that theories that are right on the lower edge of
the conformal window (and most likely below it) are the ones
with $\gamma$ approaching 1.

The twelve flavor [SU(3) fundamental] theory has been studied using FSS by a few
groups.  In \cite{DeGrand:2011cu} DeGrand obtains $\gamma$ in five
channels.  He does not combine them to a final estimate, so we
attempt to do that here.  The unweighted average over the five channels is
${\bar \gamma}=0.33$.  The r.m.s.~average jackknife error is $0.24$.  The
standard deviation of the best fit values over the five channels is $0.07$.
Combining these two sources of error in quadrature gives an overall
error of $0.25$, for a final estimate $\gamma = 0.33 \pm 0.25$.
FSS with scaling violations were considered in \cite{Cheng:2013xha};
we will discuss this paper below in the section on scaling violations.
The recent study \cite{Lombardo:2014pda}
of the twelve flavor theory obtains highly consistent results with
finite size scaling.  They also include a correction to
scaling with exponent $\omega$, which they estimate
as $\omega\sim 0.23$; cf.~their Fig.~13.  All of the results
for $\gamma$ have been summarized in Table \ref{twelvetab}.

\begin{table}
\begin{center}
\begin{tabular}{|c|c|} \hline \hline
Method & $\gamma$ \\ \hline \hline
FSS \cite{DeGrand:2011cu} & $0.33 \pm 0.25$ \\ \hline
Mode number \cite{Cheng:2011ic} & $0.61 \pm 0.05$ \\ \hline
Hyperscaling \cite{Aoki:2012eq} & 0.4 to 0.5 \\ \hline
Z factor \cite{Itou:2013kaa} & 0.044 ${}_{-0.024}^{+0.025}$ (stat.) ${}_{-0.032}^{+0.057}$ (syst.) \\ \hline
FSS w/~scale viol.~\cite{Cheng:2013xha} & 0.235(15) \\ \hline
FSS w/~scale viol.~\cite{Lombardo:2014pda} & 0.235(46) \\ \hline
Z factor \cite{Itou:2014ota} & 0.081 $\pm$ 0.018 (stat.) ${}_{-0}^{+0.025}$ (syst.) \\ \hline
Mode number \cite{Itou:2014ota} & $0.05 \leq \gamma \leq 0.08$ \\ \hline
4-loop \cite{Chetyrkin:1997dh,Vermaseren:1997fq} & 0.25 \\ \hline \hline
\end{tabular}
\caption{Summary of twelve flavor estimates of the anomalous mass dimension.  For the first
entry, corresponding to \cite{DeGrand:2011cu}, we have combined various results and errors
as described in the text. \label{twelvetab}}
\end{center}
\end{table}

\subsection{Forms of the scaling function}
Here we consider the behavior of the function $f(x)$ appearing in \myref{sfg}, where
$x=m^{1/y_m} L$ is the scaling variable.\footnote{Note that we have
returned to the notation of \myref{sfg}, since it is convenient
in the present context.  This differs from the choice \myref{scalaw} in
the previous subsection, where $x=m L^{y_m}$.}  Asymptotically, it is natural to expect
$f(x) \sim x^\lambda$; i.e., there is a leading power in the limit $x \to 0^+$.
According to the analysis of \cite{Fisher71} there are three basic cases that can occur.
The first is that $f(x)$ is a pure power:  $f(x) = A x^\lambda$.  This is unlikely to
be the case in a theory with interactions.  The second is the so-called simple case,
$f(x) = x^\lambda f_0(x)$ where $f_0(x) = f_0 + f_1 x + f_2 x^2 + \cdots$, i.e.,
$f_0(x)$ is analytic about $x = 0$.  The third is the so-called complex case,
$f(x) = x^\lambda f_0(x)$ where $f_0(x)$ is nonanalytic at $x=0$.  In this third
case, Fisher distinguishes two subcases.  The first he calls ``coincident weaker
singularities,'' where $| \ln f_0(x) | \to |\ln f_0| < \infty$ as $x \to 0$.
As an example he gives $f_0(x) = f_0 + f_1 x^\mu + \cdots$ where $0 < \mu < 1$.
The second he calls ``divergent coincident singularities,'' where 
$| \ln f_0(x) | \to \infty$ as $x \to 0$.  As examples he gives 
$f_0(x) = ( \ln x^{-1} )^\mu f_{00}(x)$ or $f_0(x) = \exp [ \nu ( \ln x^{-1} )^\mu ] f_{00}$
[we have corrected what seems to be a typo in the second example].

So which type of behavior do we have in the gauge field theories that we are studying?
Normally we think that it is necessary to take the thermodynamic limit $L \to \infty$
in order to develop thermodynamic singularities, and in particular for $\xi_L / L$ to
become infinite.  However, $x \to 0$ corresponds to holding $L$ finite while
taking $m \to 0$.  Fisher seems to imagine that this limit may yet lead to a singularity.
In a quantum field theory there is an infinite number of degrees of freedom
even at finite $L$, if we send the UV cutoff to infinity, i.e., $a \to 0$.  Of
course we have to renormalize and in a renormalizable theory this effectively
replaces the contribution of the UV modes by finite quantities, so that it is as if we have a finite
number of degrees of freedom when momentum is quantized due to finite $L$.
We thus conclude that it is unlikely that we would realize the case of divergent
coincident singularities.  However, all of the other cases Fisher cites are open to us.
In particular both the simple case and the coincident weaker singularities are
distinct possibilities.

\subsection{Corrections to scaling}
\sss{Lattice spacing}
The general scaling law for masses and decay constants is [after a slight
redefinition of Eq.~\myref{sfg}]\footnote{This just corresponds to $f(m^{1/y_m} L)
= f((m L^{y_m})^{1/y_m}) \equiv {\tilde f}(m L^{y_m})$ and then dropping the tilde.}
\beq
\xi_L/L = f(m L^{y_m})
\label{scalaw}
\eeq
Often we simplify the notation to $x=m L^{y_m}$ and write $f(x)$.
Corrections to scaling, or scaling violations, have additional terms,
which at leading order take the form
\beq
\xi_L/L = f(m L^{y_m}) + L^{-\omega} g(m L^{y_m})
\label{scavio}
\eeq
[Alternatively, see \myref{rother} below.]
Scaling violations have been incorporated into finite size scaling in the recent work \cite{Cheng:2013xha}.
This was used to resolve apparent discrepancies between the scaling exponent $\gamma$ according
to different observables in the twelve fundamental flavor theory with gauge group $SU(3)$, which has
been the source of much controversy.

Here we show that scaling violations and discretization errors may in some cases be the same thing.
Corrections to scaling are not just a feature of continuum theories.
Eq.~\myref{scalaw} is written in lattice units.  Restoring the lattice
spacing, one has
\beq
\xi_L/L = f((ma) (L/a)^{y_m})
\eeq
Now suppose there is discretization error that vanishes in the $a\to 0$ limit.
Since we can only add dimensionless quantities to the above
equation (the left-hand side is dimensionless), the discretization
error can only enter through the combinations $ma$ and $a/L$.
Thus the most general form of correction to scaling is 
$h(ma,a/L)$.  Defining $x = (ma) (L/a)^{y_m}$ we see from 
\beq
ma = (a/L)^{y_m} [ (ma) (L/a)^{y_m} ]
\label{maredf}
\eeq
that $ma = (a/L)^{y_m} x$ and so we can redefine the scaling correction
as
\beq
h(ma,a/L) \equiv {\tilde h}(a/L,x)
\eeq
Expanding in $a$, holding $x$ fixed, we obtain
\beq
{\tilde h}(a/L,x) = (a/L)^{\omega_1} g_1(x) + (a/L)^{\omega_2} g_2(x) + \cdots
\eeq
where $\omega_1 < \omega_2 < \cdots$.
Thus we see that \myref{scavio} is the leading order correction in powers of
the lattice spacing.

Irrelevant operators are also tied up with the scaling violations.  By RG arguments given above, if $g_i$ are the irrelevant couplings
and $y_i < 0 $ are their scaling dimensions, then
\beq
M_H/L = f(L m^{1/y_m}, g_i m^{-y_i/y_m} )
\eeq
In fact, the presence of irrelevant operators is often traced to the nonzero lattice spacing, so that
there is a connection with the scaling violations that were discussed above.
Next we specialize this to the gauge coupling, which is irrelevant in a theory with an IRFP.

\sss{Gauge coupling}
Eq.~\myref{nzg} describes infinite volume.  We can extend to $g\not=0$ (recall that $g \to g-g_*$ has been made)
RG transformation,
\beq
C_H(t;{\hat m},g,L,\mu) = \lambda^{-2\gamma_H} C_H(t;{\hat m}',g',L,\mu')
\eeq
Repeating the steps in \myref{rescale-arg}, we find that
\beq
C(t;{\hat m}',g',L,\mu') = \lambda^{-2d_H} C(\lambda^{-1} t; {\hat m}', g', \lambda^{-1} L, \mu)
\eeq
Again taking $\lambda^{-1} = {\hat m}^{1/y_m}$, we find
\beq
C(t;{\hat m},g,L,\mu) = \mu^{2 d_H} {\hat m}^{2 \Delta_H / y_m} 
F(\mu {\hat m}^{1/y_m} t, 1, {\hat m}^{-y_g/y_m} g, {\hat m}^{1/y_m} L \mu)
\eeq
It follows that the ``hadron'' mass will satisfy
\beq
M_H L = f_H(x,{\hat m}^{-y_g/y_m} g), \quad x = {\hat m}^{1/y_m} L
\eeq
In \cite{Cheng:2013xha} this is expanded about small $g$ to obtain
\beq
M_H L = F_H(x) \( 1 + G_H(x) {\hat m}^{-y_g/y_m} g + \ord{g^2} \)
\label{rother}
\eeq
They furthermore approximate $G_H(x)$ as a constant $c_G$ over the
range of $x$ that they consider.  Defining $\omega = -y_g/y_m$ and
$c_0 = c_G g$, they fit the spectrum to
\beq
\frac{M_H L}{1 + c_0 m^\omega} = F_H(x)
\eeq
allowing $c_0$ and $\omega$ to be optimized in the fit, and $F_H(x)$ as the concatenation
of two quadratic functions whose coefficients are also fit.  Finally, the $c_0$
values are allowed to be different for each value of the bare lattice coupling $\beta_F$
that multiplies the part of the gauge action in the fundamental representation.
(Their simulation also involve an adjoint gauge action with coefficient $\beta_A$.)
They are able to obtain reasonably consistent values for the exponents
when the pion, rho and pion decay constant are included in a combined fit,
although the $\chi^2$ increases by a factor of two compared to the fit with
only the pion mass.

\section{Schr\"odinger functional}
\label{schrodinger}
The Schr\"odinger functional approach simulates on the four-dimensional
cylinder $W \times {\bf T}^3$ where $W = [0,T]$ is a section of the real line
and ${\bf T}^3$ is a three-torus.  The three torus is represented
as a cubic lattice with $L/a$ sites in each direction, and
periodic boundary conditions.  The temporal extent is related
to the spatial extent by a fixed aspect ratio, $T = \zeta L$.
The boundary conditions at the boundaries
$t=0$ and $t=T$ are such that the classical solution to the field equations
corresponds to a constant $\ord{1/L}$ chromoelectric field.  A classic use of this
is to obtain the running coupling by measuring the response of the
system to this background chromoelectric field.  However, it can
also be used for determining the renormalization constant associated
with the mass operator $\psib \psi$, and hence the anomalous
mass dimension.  
Recently it has also been discussed in the context
of determining the anomalous dimension of four-fermion operators,
which may play an important role in modifying the predictions of
walking technicolor theories.

\subsection{Anomalous mass dimension}
The renormalization of the quark mass and the corresponding anomalous
dimension in the Schr\"odinger functional scheme was developed originally 
in \cite{Sint:1998iq,Capitani:1998mq,DellaMorte:2005kg}.  Its early use in 
the context of walking technicolor
was seen in \cite{Bursa:2009we}.

The method here begins with the partially conserved axial current (PCAC) relation.
This can be written
\beq
Z_A \p_\mu A_\mu^a = 2 m_{\text{ren.}} Z_P P^a
\eeq
where $A_\mu^a$ and $P^a$ are the bare axial current and pseudoscalar density.
$Z_A$ and $Z_P$ are renormalization constants, and $m_{\text{ren.}}$ is the
renormalized PCAC mass (we assume a degenerate $N_f$ flavor model).  On the other hand
we have the bare relation
\beq
\p_\mu A_\mu^a = 2 m_{\text{latt.}} P^a
\eeq
Comparing these two we see that
\beq
m_{\text{latt.}} = m_{\text{ren.}} \frac{Z_P}{Z_A}
\eeq
The quantity $m_{\text{latt.}}$ is obtained from lattice correlation functions
without any reference to the renormalization scale $\mu$, hence is $\mu$ independent.
From this we immediately obtain
\beq
\gamma_m = - \frac{d \ln m_{\text{ren.}}}{d \ln \mu} = \frac{d}{d \ln \mu} \ln \( \frac{Z_P}{Z_A} \)
\eeq
The renormalization factor $Z_A$ is conventionally determined by the
requirement that the current algebra take its canonical form.  As a result
of the nonlinearity of this set of relations, the anomalous dimension of $Z_A$ must
vanish,\footnote{This argument has been presented, for instance, in \cite{Sint:1998iq}.}
\beq
\frac{d \ln Z_A}{d \ln \mu} = 0
\eeq
The result is therefore that
\beq
\gamma_m = \frac{d \ln Z_P}{d \ln \mu}
\eeq
That is, the anomalous mass dimension can be determined from the pseudoscalar
density renormalization factor.  It is interesting that this density is related to
the scalar density $\psib \psi$, normally associated with $\gamma_m$, by a chiral
rotation.

In the Schr\"odinger functional approach, $\mu = 1/L$.  
Hence
\beq
\gamma_m = - \frac{d \ln Z_P}{d \ln L}
\eeq

We must now explain how $Z_P$ is obtained.  One introduces a notation for the boundary fermions
\beq
&& \zeta({\bf x}) = q(0,{\bf x}), \quad \zeta'({\bf x}) = q(T-1,{\bf x}),
\ddd {\bar \zeta}({\bf x}) = {\bar q}(0,{\bf x}), \quad {\bar \zeta}'({\bf x}) = {\bar q}(T-1,{\bf x})
\label{bfs}
\eeq
Adopting the notation of \cite{Sint:1998iq}, one then forms correlation functions involving
the boundary fields:
\beq
f_P(x_0) &=& - \frac{1}{3} \sum_{{\bf y},{\bf z}} \langle P^a(x) {\bar \zeta}({\bf y}) \gamma_5
\half \tau^a \zeta({\bf z}) \rangle \nnn
f_1 &=& - \frac{1}{3L^6} \sum_{{\bf u},{\bf v},{\bf y},{\bf z}} \langle {\bar \zeta}'({\bf u})
\gamma_5 \half \tau^a \zeta'({\bf v}) {\bar \zeta}({\bf y}) \gamma_5 \half \tau^a \zeta({\bf z}) \rangle
\label{ffcor}
\eeq
The renormalized versions of these are
\beq
f_P^{\text{ren.}} = Z_P Z_\zeta^{-1} f_P, \quad
f_1^{\text{ren.}} = Z_\zeta^{-2} f_1
\eeq
where $Z_\zeta$ is the wavefunction renormalization of the $\zeta, \zeta'$ boundary
fields, e.g., $\zeta = \sqrt{Z_\zeta} \zeta_{\text{ren.}}$.  Then it is easy to see that
\beq
Z_P = \frac{f_P^{\text{ren.}}(x_0)}{\sqrt{f_1^{\text{ren.}}}} \frac{\sqrt{f_1}}{f_P(x_0)}
\eeq
One typically chooses $x_0=T/2$.  To a certain extent, the renormalized values 
$f_P^{\text{ren.}}(T/2)$ and $f_1^{\text{ren.}}$ are arbitrary, so the convention
is to take these equal to the tree-level values
\beq
\frac{f_P^{\text{ren.}}(T/2)}{\sqrt{f_1^{\text{ren.}}}} \equiv \frac{f_P^{\text{tree}}(T/2)}{\sqrt{f_1^{\text{tree}}}}
\equiv c
\label{renrat}
\eeq
Then
\beq
Z_P = c \frac{\sqrt{f_1}}{f_P(T/2)}
\eeq
The constant $c$ evaluates to
\beq
c = \sqrt{N_c} + \ord{a^2}
\eeq
so that it is $L$ independent.  Thus we arrive at a somewhat strange situation where
the renormalized ratio in \myref{renrat} is independent of the RG scale $\mu=1/L$
whereas the bare ratio $\sqrt{f_1}/f_P(T/2)$ carries all of the $\mu=1/L$ dependence.
This is a consequence of choosing the scheme \myref{renrat} and identifying $\mu$
with the inverse size of the lattice $1/L$.

It is interesting to contrast this with what would happen with periodic
boundary conditions with $t$ now ranging $t = 0, a, 2a, \ldots, 2L-a$.
One could retain the ``boundary'' fields \myref{bfs} but with $T \to L$, which
are really just fields on the timeslice at the origin, and the midpoint
timeslice.  However now the correlation function \myref{ffcor} have
the dependence on $L$ of (they involve pseudoscalar operators, and
are hence dominated by the pion):
\beq
f_P(x_0) \sim e^{-m_\pi x_0} + e^{-m_\pi (2L - x_0)}, \quad
\sqrt{f_1} \sim e^{-m_\pi L / 2}
\eeq
Forming the relevant ratio
\beq
\frac{f_P(x_0)}{\sqrt{f_1}} \sim e^{-m_\pi (x_0-L/2)} + e^{-m_\pi (3L/2 - x_0)}
\eeq
we see that there is no choice of $x_0$ which would cancel the $L$ dependence.
The result is that the constant $c$ which would appear in
\beq
Z_P = c \frac{\sqrt{f_1}}{f_P(x_0)}
\eeq
would not be independent of $L$, and it would need to be known in order to
obtain the anomalous mass dimension.  This shows the clear superiority of
the Schr\"odinger functional approach.

As an example of this method, in \cite{DeGrand:2012yq} the anomalous mass dimension was obtained
for sextet ``QCD.''  
Because of a slow running that was observed, they were able to fit
to 
\beq
Z_P(L) = Z_P(L_0) \( \frac{L_0}{L} \)^\gamma
\eeq
to obtain $\gamma$.  One sees from their analysis that $\gamma \lappeq 0.4$
for the action that is able to probe the strong coupling regime where
the fixed point is supposed to exist.
This approach has also been used for SU(2) with two flavors in the adjoint
representation \cite{Bursa:2009we,DeGrand:2011qd}.  The results have been
presented above in Table \ref{taball}.
Another study of the anomalous mass dimension using the Schr\"odinger functional
technique is \cite{Hayakawa:2013yfa}.  There they study SU(2) gauge theory with
six flavors of fundamental representation fermions. They obtain $0.26 \lappeq \gamma \lappeq 0.74$.
In \cite{DeGrand:2012qa}, SU(4) gauge theory with decuplet fermions was studied,
yielding $\gamma \sim 0.4$, quite similar to what was seen for SU(2) and SU(3)
two-index symmetric representation fermions.

\subsection{Four-fermion operators}
Ref.~\cite{Debbio:2014vea} takes some preliminary steps toward using the Schr\"odinger
functional to determine the anomalous dimensions of four-fermion operators.  This is
a very important direction to explore, because one would like to obtain the full
spectrum of critical indices in an interacting four-dimensional conformal field theory.
An alternative method would be to include the four-fermion operator in the lattice
action, and then use finite size scaling with the coefficient of that operator.\footnote{Two
problems would arise: (1) additive renormalization would require a subtraction to find
the scaling variable; (2) in many cases a sign problem would result.}

\section{The eigenmode number approach}
\label{sec:modenumber}
\subsection{Early steps}
In \cite{DeGrand:2009et} it was shown that the scaling \myref{condscaling} implies that
the density of Dirac eigenvalues $\rho(\lambda)$ has the behavior
\beq
\rho(\lambda) \sim \lambda^{\eta_{\psib \psi}}
\eeq
This was rederived in \cite{DelDebbio:2010ze}.
Hence we can measure the spectrum of $D$ and obtain an estimate of $\gamma_*$.

\subsection{The mode number development}
Related to the eigenvalue distribution $\rho(\lambda)$, one can define a
quantity known as the ``mode number'' which has a number of useful properties.
In \cite{Patella:2012da} a precise result was achieved, $\gamma_m = 0.371(20)$
for SU(2) with two Dirac flavors in the adjoint representation.  
Other recent results using this method include
\cite{Cheng:2011ic,Hasenfratz:2012fp,deForcrand:2012vh,Cheng:2013eu,Cheng:2013bca,Landa-Marban:2013oia,
DelDebbio:2013hha,Cichy:2013eoa,Perez:2015yna,Keegan:2015cba}.
For instance in \cite{Cheng:2011ic} it was found in the SU(3) twelve flavor theory
that $\gamma = 0.61(5)$.
The integral of the spectral
density $\rho(\lambda)$ of the Dirac operator (a.k.a.~the mode number $\nu(\lambda_{\text{max}})$) 
is used to predict the anomalous dimension of the mass
operator.  On the lattice with Wilson type fermions the spectrum is complex, so what
one actually looks at is
\beq
D_L^\dagger D_L \psi_\lambda = |\lambda|^2 \psi_\lambda
\eeq
so that it is really $\rho(|\lambda|)$ which is analyzed.  Some analyses also
use eigenvalues $\mu^2$ of the massive operator $D_L^\dagger D_L + m^2$, in which case
eigenvalues are obtained as $|\lambda|^2 = \mu^2 - m^2$.  For small values of $|\lambda|$
the eigenvalues are almost imaginary, so this becomes equivalent to the Euclidean
continuum spectral density
\beq
D_C \psi_\lambda = i \lambda \psi_\lambda
\eeq
$\rho=\rho(\lambda)$ up to lattice artifacts and an obvious double-counting related
to the fact that for every nonzero eigenvalue $i \lambda$ there is a corresponding
eigenvalue $i \lambda' = -i \lambda$.  From this point on we will simply write
$\lambda$ rather than $|\lambda|$, ignoring this lattice detail.

From RG arguments, reviewed above, we know that
\beq
\rho(\lambda) \sim \lambda^{(3-\gamma)/(1+\gamma)}
\eeq
The mode number is then determined from
\beq
\nu(\lambda_{\text{max}}) = V \int_0^{\lambda_{\text{max}}} d\lambda ~ \rho(\lambda)
\eeq
where $V$ is the volume of space.  If the massive Dirac operator is used, then
the mode number also depends on this quantity:  $\nu = \nu(\lambda_{\text{max}},m)$.
The chiral condensate, in the case of spontaneous chiral symmetry breaking,
can be obtained from the mode number according to the formula \cite{Giusti:2008vb}
\beq
\Sigma = \frac{\pi}{2V} \frac{d \nu}{d \lambda_{\text{max}}}
\eeq

One of the advantages to using the mode number is that it is known to be renormalization
group invariant \cite{DelDebbio:2005qa,Giusti:2008vb}.  
In addition, the eigenvalues are only multiplicatively
renormalized.  Hence, we avoid the divergent ($\ord{1/a^3}$ at finite lattice spacing)
subtractions that would be necessary if we were to use the condensate relation \myref{condscaling}.
Note that the multiplicative renormalization is a non-issue because if $\lambda = Z \lambda_0$
relates the renormalized and bare eigenvalue, we still have
\beq
\rho(\lambda_0) \sim \lambda_0^{(3-\gamma)/(1+\gamma)}
\eeq
All that has happened is that the constant of proportionality has been modified.
Another advantage of the mode number method that should be emphasized is that it
allows the determination of $\gamma$ from a single lattice simulation.  By contrast,
the finite size scaling approach requires many $L, m$ values to be simulated (although
many of these can be small $L$ and larger $m$, so the actual computational
cost is not that high).  However, the most significant benefit seems to be accuracy.

In practice, a finite mass should be simulated in order to avoid large finite
volume effects.  In addition, the anomalous dimension that we are interested in
is a property of the IR of the theory.  Thus one must find an optimal window
$\lambda_1 < \lambda < \lambda_2$ where the fit should be applied.  How does
one determine this window?  $\lambda_2$ can be identified because of asymptotic
freedom:  at large $\lambda$ the anomalous dimension vanishes and $\rho \sim \lambda^3$.
So one avoids that region by taking $\lambda_2$ sufficiently small.  $\lambda_1$ can
be identified from the fact that it is mass-dependent.  Looking at the scaling of
$\rho$ at small $\lambda$, one looks for the region that changes its behavior
significantly as the mass is adjusted.  Clearly there is some uncertainty
in choosing the optimum values of $\lambda_{1,2}$, but this is no different
than any other method---we characterize this a systematic uncertainty.
The encouraging fact found in \cite{Patella:2012da} is that the PCAC mass
does not need to be very small in order to open up a reasonable window
and obtain accurate results for $\gamma$.  Indeed other groups have also
applied this method with much success.

One powerful technique can be applied here which is a stochastic determination of the
mode number
\beq
\nu(\Omega) = \frac{1}{V} \vev{ \tr \mathbb{P} }
\eeq
where $\mathbb{P}$ is a projector
\beq
\mathbb{P} = h^4 \( 1 - \frac{2 \Omega_*^2}{D^\dagger D + \Omega_*^2} \)
\eeq
where $h(x)$ is a polynomial that approximates $\theta(-x)$ within a particular
range of $|x|$.  
The traces of inverses are computed by solving (e.g., using conjugate gradient)
\beq
(D^\dagger D + \Omega_*^2)^n(x,y) X_i(y) = \eta_i(x)
\eeq
with $N_r$ random sources $\eta_i$.  Then
\beq
\tr (D^\dagger D + \Omega_*^2)^{-n} = \frac{1}{N_r V} \sum_{i,x} \tr_{s.c.} ( X_i(x) \eta_i^*(x) ) + \ord{1/\sqrt{N_r}}
\eeq
where the trace on the r.h.s.~is only over spin and color indices.  Typically one fully dilutes
over spin-color indices (factor of 12) and $N_r=\ord{10^2}-\ord{10^3}$ may be required for an accurate estimate.\footnote{Dilution
may be of some help here, although it is not clear because the trace is a short-distance quantity.}
These inversions must be performed on each gauge field configuration.
This is a perfect workload for GPUs, which optimally perform a large number of inversions.
It would also entail a small project to modify the GPU code (probably QUDA) to implement
$(D^\dagger D + \Omega_*^2)^n$ inside the inverter.
Note also that we only need to perform the inversions for the largest power of $n$
appearing in $h^4(x)$, since traces of lower powers can be obtained from
\beq
\tr (D^\dagger D + \Omega_*^2)^{-(n-m)} = \frac{1}{N_r V} \sum_{i,x,y}
\tr_{s.c.} ( (D^\dagger D + \Omega_*^2)^m(x,y) X_i(y) \eta_i^*(x) )
\eeq

The lattice data is fit to
\beq
a^{-4} \nu(\Omega) = a^{-4} \nu_0 + A [ (a\Omega)^2 - (am)^2 ]^{2/(1+\gamma)}
\eeq
Here, $m = Z_A m_{\text{PCAC}}$ where $m_{\text{PCAC}}$ is the bare PCAC mass.  It is amusing
that since $m_{\text{PCAC}}$ is known and $m$ is obtained from the fit, the mode
number analysis provides a way to obtain the renormalization constant $Z_A$.

The mode number analysis has been extended to address a scale dependent anomalous
mass dimension $\gamma(\mu)$ in \cite{Cheng:2013eu}.  This is a method that can
be applied both in theories with a conformal fixed point, and in theories that
are confining in the infrared.  In \cite{Cheng:2013eu} it is shown how to
combine lattices of different volumes and bare couplings in order to obtain a more
complete picture.  This provides added confidence in the approach because they
are able to follow the dynamics from asymptotic freedom in the ultraviolet to
spontaneous chiral symmetry breaking in the infrared in the context of the $N_f=4$
theory (SU(3) with fundamental flavors).

In this study, the matching of different $\beta$ was done with the following rescaling
of the eigenvalues $\lambda$ for $\gamma < 2$:
\beq
\lambda' = \( \frac{a}{a'} \)^{1+\gamma} \lambda
\eeq
This is based on the fact that the eigenvalues should scale like the
mass parameter.  For the infrared where couplings
are larger and $\gamma > 2$, the prescription in \cite{Cheng:2013eu} is to terminate the above scaling
at $\gamma=2$ and replace it with
\beq
\lambda' = \( \frac{a}{a'} \)^{2} \lambda
\eeq
for $\gamma > 2$.  Empirically, this causes the curves to fall on top of
each other giving a fairly smooth overall curve.

Recently, in an effort to better understand the systematic
uncertainties associated with the mode number approach,
Keegan has studied the Schwinger model, where analytic
results are also available \cite{Keegan:2015cba}.
  This paper considers
the $N_f=0,2$ models with small fermion mass and follows
an approach similar to \cite{Cheng:2013eu} in that it connects
a range of $\beta$ values by rescaling the eigenvalues
by a power of the lattice spacing.  By doing this, Keegan
is able to go from very small eigenvalues to very large
eigenvalues and follow the flow of $\gamma(\lambda)$.
Indeed, in the $N_f=2$ model he sees that in the IR $\gamma \approx 0.5$,
consistent with the analytic prediction, and that in the UV $\gamma \approx 0$,
also consistent with the analytic prediction (asymptotic freedom).
In addition he uses the spectral density $\rho(\lambda)$ directly
and finds a consistent result, though with much larger statistical
errors.  This work clears up the contradiction found in \cite{Landa-Marban:2013oia},
which only used one value of $\beta$.

Ref.~\cite{Athenodorou:2014eua}
uses the mode number to find $\gamma \sim 0.9$ for $N_f=1$ SU(2) adjoint,
consistent with what they found using finite size scaling.  This rather
large value is another indication that theories at the lower edge
of the conformal window are the ones that are the most likely to
be phenomenologically viable in terms of the size of $\gamma$.
Another interesting development is the recent paper \cite{Perez:2015yna},
which used Eguchi-Kawai reduction with the mode number to estimate $\gamma$ in the large
$N_c$ limit.

\section{Conclusions}
Because of its phenomenological importance, the anomalous mass dimension in conformal and
nearly conformal theories is the quantity that has been most studied on the lattice in the
last few years.  As has been seen in this review, a number of techniques have been developed
for obtaining fairly accurate estimates.  Additional exponents, such as the correction
to scaling index, are beginning to also be included.  It is hoped that the coming years will
see the computation of other critical exponents in conformal field theories on the lattice.

We have learned various facts about the anomalous mass dimension of phenomenological importance.
It seems that it is the theories that are on the lower edge of the conformal window which
have $\gamma \approx 1$, and in other conformal theories $\gamma$ is too small.  Four fermion
operators coming from extended technicolor have been suggested as a way to push theories
out of the conformal window and increase $\gamma$, as is needed for a viable walking
technicolor theory \cite{Fukano:2010yv}.  Alternatively, giving mass to some subset of
the fermions may also produce the same effect \cite{Brower:2014ita}.  In both cases it would be very
useful to have careful studies of the anomalous mass dimension from the lattice.

\section*{Acknowledgements}
The author was supported in part by the Department of Energy, Office of Science, Office of High Energy Physics,
Grant Nos.~DE-FG02-08ER41575 and DE-SC0013496.

\bibliography{anomdimrvw}
\bibliographystyle{JHEP}

\end{document}